\documentclass[twocolumn,apj,numberedappendix]{openjournal}

\usepackage[dvipsnames, table]{xcolor}
\definecolor{PineGreen}{HTML}{01796F}
\definecolor{Cerulean}{HTML}{007BA7}

\usepackage{xcolor}
\usepackage{textgreek}
\usepackage{enumitem}
\usepackage[utf8]{inputenc}
\usepackage[english]{babel}

\usepackage{hyperref}
\hypersetup{
    unicode, 
    colorlinks=true,
    linkcolor=linkcolor,
    citecolor=linkcolor,
    filecolor=linkcolor,
    urlcolor=linkcolor,
}
\usepackage{color,colortbl}
\definecolor{linkcolor}{rgb}{0.0,0.3,0.5}
\usepackage{tensind}
\tensordelimiter{?}
\DeclareGraphicsExtensions{.bmp,.png,.jpg,.pdf}
\usepackage{verbatim}
\usepackage[normalem]{ulem}
\usepackage{orcidlink}
\usepackage{soul}
\usepackage{multirow}

\usepackage[style=list]{glossaries}

\graphicspath{ {./figs/} }

\usepackage{newtxtext,newtxmath}

\usepackage[T1]{fontenc}

\usepackage{graphicx}	
\usepackage{amsmath, amssymb, graphicx, color}

\usepackage{xspace}

\makeglossaries
\glsdisablehyper
\newacronym{VMS}{VMS}{very massive star}
\newacronym{HD}{HD}{Humphreys-Davidson}
\newacronym{HR}{HR}{Hertzsprung-Russell}

\newacronym{MS}{MS}{main sequence}
\newacronym{ZAMS}{ZAMS}{zero-age main sequence}
\newacronym{TAMS}{TAMS}{terminal-age main sequence}
\newacronym{HG}{HG}{Hertzprung-gap}
\newacronym{CHeB}{CHeB}{core-helium burning}
\newacronym{AGB}{AGB}{asymptotic giant branch}
\newacronym{RGB}{RGB}{red giant branch}
\newacronym{BSG}{BSG}{blue supergiant}
\newacronym{YSG}{YSG}{yellow supergiant}
\newacronym{RSG}{RSG}{red supergiant}
\newacronym{WR}{WR}{Wolf-Rayet}
\newacronym{LBV}{LBV}{luminous blue variable}
\newacronym{BH}{BH}{black hole}

\newacronym{RLOF}{RLOF}{Roche-lobe overflow}

\newacronym{PPSN}{PPSN}{pair-instability pulsation supernova}
\newacronym{PSN}{PSN}{pair-instability supernova}

\newacronym{LVK}{LVK}{LIGO--Virgo--KAGRA}
\newacronym{GW}{GW}{gravitational wave}

\newacronym{MW}{MW}{Milky Way}
\newacronym{LMC}{LMC}{Large Magellanic Cloud}
\newacronym{SMC}{SMC}{Small Magellanic Cloud}

\newacronym{Mdot}{$\dot M$}{mass-loss rates}

\newacronym{LLM}{LLM}{Large Language Model}

\definecolor{burgundy}{rgb}{0.5, 0.0, 0.13}
\definecolor{coral}{rgb}{1.0, 0.5, 0.31}

\newcommand{\ST}[1]{{\color{coral}\textbf{[Paper II]}}}

\newcommand{\Msun}{\ensuremath{\,\rm{M}_{\odot}}\xspace}

\newcommand{\UCSD}{Department of Astronomy and Astrophysics, University of California, San Diego, La Jolla, CA 92093, USA}
\newcommand{\ITA}{Universit\"at Heidelberg, Zentrum f\"ur Astronomie (ZAH), Institut f\"ur Theoretische Astrophysik, Albert Ueberle Str. 2, 69120, Heidelberg, Germany}
\newcommand{\Padua}{Dipartimento di Fisica e Astronomia Galileo Galilei, Università di Padova, Vicolo dell’Osservatorio 3, I–35122 Padova, Italy}
\newcommand{\CAMK}{Nicolaus Copernicus Astronomical Center, Polish Academy of Sciences, ul. Bartycka 18, 00-716 Warsaw, Poland}
\newcommand{\PragueAstro}{Astronomický ústav, Akademie věd Ceské republiky, Fričova 298, 251 65 Ondřejov, Czech Republic}
\newcommand{\IAASARS}{IAASARS, National Observatory of Athens, 15236 Penteli, Greece}
\newcommand{\UAthens}{National and Kapodistrian University of Athens, 15784 Athens, Greece}

\usepackage{xspace}

\defcitealias{Romagnolo_2026_AtlasI}{Atlas I}
\newcommand{\AtlasFirst}{\citet{Romagnolo_2026_AtlasI} (hereafter \citetalias{Romagnolo_2026_AtlasI})}
\newcommand{\AtlasI}{\citetalias{Romagnolo_2026_AtlasI}\xspace}
\newcommand{\AtlasIp}{\citepalias{Romagnolo_2026_AtlasI}\xspace}

\begin{document}

\title{The Stellar Winds Atlas II: Black Hole Formation at Solar Metallicity}

\author{
    Amedeo Romagnolo\orcidlink{0000-0001-9583-4339}$^{1,2,3,4}$,
    Floor S. Broekgaarden\orcidlink{0000-0002-4421-4962}$^{3}$,
    Alex C. Gormaz-Matamala\orcidlink{0000-0002-2588-2391}$^{5}$,
    Lucas M. de Sá\orcidlink{0000-0003-3109-9042}$^{1}$,
    Daniel Pauli\orcidlink{0000-0002-5453-2788}$^{6}$,
    Avishai Gilkis\orcidlink{0000-0001-8949-5131}$^{7}$,
    Lumen Boco\orcidlink{0000-0003-3127-922X}$^{1}$,
    Michela Mapelli\orcidlink{0000-0001-8799-2548}$^{1,2,8,9}$,
    Konstantinos Antoniadis\orcidlink{0000-0002-3454-7958}$^{10,11,12}$, and
    Laya Binu\orcidlink{0009-0009-5823-4399}$^{3}$
}
\email{amedeoromagnolo@gmail.com}

\affiliation{
    $^1$\ITA \\
    $^2$\Padua \\
    $^3$\UCSD \\
    $^4$\CAMK \\
    $^5$\PragueAstro \\
    $^6$Institute of Astronomy, KU Leuven, Celestijnenlaan 200D, 3001 Leuven, Belgium\\
    $^7$Institute of Astronomy, University of Cambridge, Madingley Road, Cambridge CB3 0HA, United Kingdom \\
    $^8$ Universität Heidelberg, Interdiszipliäres Zentrum für Wissenschaftliches Rechnen, D-69120 Heidelberg, Germany\\
    $^9$ INFN, Sezione di Padova, Via Marzolo 8, I–35131 Padova, Italy\\
    $^{10}$International Hellenic University, 62124 Serres, Greece \\
    $^{11}$\IAASARS \\
    $^{12}$\UAthens
}



\begin{abstract}

Stellar winds are a primary source of uncertainty in predicting the masses of black holes (BHs) from massive stars. At solar metallicity, theoretical models lead to widely divergent results due to differing wind prescriptions. A key obstacle remains the lack of systematic investigations across a common parameter space. To address this, we construct a ``Wind Atlas'' using detailed 1D MESA stellar evolution models and population synthesis techniques to estimate the Galactic population of solar metallicity BH progenitors. We systematically investigate 14 distinct wind models, ranging from the most traditional and widespread prescriptions to the most recent. By evaluating stellar evolution across this extensive grid, we show that the final BH mass is dictated by a fundamental bifurcation: whether a star collapses as a cool supergiant or is first stripped of its envelope to become a Wolf-Rayet (WR) star. If a star enters the WR stage, its strong thick winds dominate, making the final mass sensitive to the WR wind prescription while largely erasing the memory of its prior mass-loss history. Conversely, stars that face core collapse as supergiants form significantly more massive BHs, producing a mass peak around an initial mass of 40~$M_\odot$. Rather than simply reproducing these divergent outcomes, our comprehensive evaluation demonstrates that this bifurcation is universally controlled by the highly uncertain mass loss during the cool supergiant phase. This framework strongly constrains the problem of BH mass prediction by identifying two key bottlenecks for future studies: envelope stripping efficiency and WR mass-loss rates. Our atlas provides a clear baseline for interpreting current theoretical discrepancies and testing wind models against observational constraints, such as the Galactic WR/OB population ratio.
\end{abstract}


\maketitle

\section{Introduction}
\label{sec:intro}

\Glspl{BH} are one of the end products of massive stars, but the vast majority of the \gls{BH} population remains unobserved. The Milky Way alone is estimated to host $\sim 10^8-10^9$ stellar-mass \glspl{BH} \citep{vandenHeuvel_1992, BrownBethe_1994, Samland_1998, Olejak_2020a}, but so far only $\sim$30 \citep{Casares_2014,BlackCAT,Khokhlov_2018,Thompson_2019,Saracino_2021,ElBadry_2022,ElBadry_2023,GaiaBH3} have been detected via electromagnetic observations.
In contrast, the \gls{LVK} collaboration has observed over two hundred \gls{BH} merger events \citep{GWTC1,GWTC2,GWTC3,GWTC4,GWTC5}, with the new-generation observatories such as the Einstein Telescope and Cosmic Explorer predicted to detect more than one \gls{BH} merger per day \citep{Branchesi_2023, CosmicExplorer_2023,EinsteinTelescope_BlueBook_2026}.

The masses of the \gls{LVK} \glspl{BH} frequently exceed those found via electromagnetic detections \citep{GWTC5}. Because the final mass of a stellar remnant is fundamentally governed by the mass-loss history of its progenitor, this discrepancy points toward variations in stellar wind efficiencies. A leading interpretation is that these massive mergers originate from low-metallicity environments, which are more common at high redshifts \footnote{Connecting these redshift-dependent merger properties to their massive star progenitors remains however highly sensitive to the assumed cosmic star formation and chemical enrichment histories \citep{Boco_2026,Levina_2026}}. In these regimes, reduced stellar winds enable stars to retain more mass, while simultaneously leading to more compact progenitor radii. This combination allows isolated binaries to survive interactions and collapse into heavier remnants, provided their cores avoid pair-production instabilities such as \gls{PPSN} and \gls{PSN} \citep{Chruslinska_2019b,Boco_2021}\footnote{Alternatively, these massive gravitational wave sources could come from second-generation \citep[e.g.][]{Gerosa_2017} or primordial \gls{BH} \citep{Carr_2016,Sasaki_2018} mergers.}.

Beyond setting the final mass of compact remnants, mass loss has a crucial impact on the evolution of massive stars ($M_{\rm ZAMS}\gtrsim~8~M_\odot$) \citep{Meynet_1994,Eggenberger_2021,Garcia_2025} throughout every evolutionary phase. By altering the structural evolution, winds explicitly dictate a star's response to mass transfer events \citep[e.g.][]{Pavlovskii_2017,Marchant_2021,Klencki_2020,Romagnolo_2025,Klencki_2026} and shape the overarching \gls{BH} mass distribution across different populations \citep[e.g.][]{Merritt_2025,vanSon_2025}. Furthermore, this mass loss provides essential mechanical feedback \citep{Ramachandran_2018,Ramachandran_2019,Sander_2020} and local chemical enrichment to the surrounding environment \citep{Maeder_1983,Dray_2003,Dray_2003b,Farmer_2021}


Stellar winds at solar metallicity ($Z_\odot$) provide a crucial, observationally-rich testbed for stellar evolution theory. At this metallicity, mass loss is highly pronounced, and the Milky Way offers a wealth of detected high-mass stars \cite[e.g.][]{Martins_2005,Crowther_2010,Clark_2012} and stellar-mass \glspl{BH} in binaries that can be used for calibrations, like the well studied case Cygnus X-1 \citep{Webster_1972,MillerJones_2021,Ramachandran_2025}. Consequently, many studies have presented estimates for the final masses of \gls{BH} progenitors at $Z_\odot$ with varying assumptions for stellar winds \citep[e.g.][]{Belczynski_2010,Giacobbo_2018,Mapelli_2020,Bavera_2023,Martinet_2023,Gilkis_2024,Kruckow_2024, Romagnolo_2024,Vink_2024,Costa_2025,Hirschi_2025,Ugolini_2025,vanSon_2025}. However, these models yield widely divergent predictions, sparking debates on the underlying physics.
A key obstacle is the lack of systematic investigations across a common parameter space, with most studies offering qualitative explanations for their results rather than a direct comparison of their assumptions against others.

Building upon recent work that quantifies the theoretical uncertainties of wind-driven mass loss through both analytical calculations and detailed stellar evolution models \AtlasFirst, we address this lack of cross-model investigations in stellar winds recipes by constructing a Wind Atlas for massive stars at $Z_\odot$. Our goal is to systematically connect and contextualize the wide range of theoretical predictions by exploring an extensive set of stellar wind prescriptions, including new models not previously incorporated into evolutionary calculations. We analyze how different wind implementations lead to different mass-loss histories and \gls{BH} masses, focusing not only on whether winds are stronger or weaker, but on how their underlying assumptions reshape a star's entire life cycle, with a major focus on the time spent in different wind phases. By comparing the resulting evolutionary tracks and \gls{BH} masses with modern observational catalogs \citep{Martins_2008,Pauli_2025,GormazMatamala_2025} of massive stars in the Milky Way, we provide a framework to interpret current discrepancies and deliver a procedure to test stellar wind models against future empirical results.

\section{Method} 
\label{sec:Method}

The evolution of a massive star is affected by several distinct phases of wind-driven mass loss. In standard theoretical frameworks, these phases are categorized into the line-driven regimes of hot stars, which include both optically thin and optically thick winds, and the separate regimes governing cool supergiants. While non-cool supergiants, such as blue supergiants, possess hot atmospheres where line driving remains the dominant mechanism, cool supergiants, which encompass yellow and red supergiants, experience a physical transition away from line driving. We specifically choose the term cool supergiant winds over dust-driven winds because the primary mass-loss driver in this low-temperature regime remains highly uncertain. Although dust opacity can reinforce outflows in the coolest red supergiants, recent models suggest that alternative mechanisms, such as atmospheric turbulent pressure or convective boil-off shocks, may initiate the wind, rendering dust a consequence rather than the root cause. A comprehensive analysis of the underlying physics, a comparison of competing theoretical formulations, and a discussion of key challenges such as wind clumping, the bistability jump, and the criteria for transitioning to a \gls{WR} star are presented in our companion paper \AtlasI. In this study we analyze the uncertainties in stellar winds prescriptions with a large range of stellar evolution models for $Z_\odot$ massive stars.

\subsection{{\tt MESA} Stellar Evolution setup}

We use version 24.08.1 of Modules for Experiments in Stellar Astrophysics~\citep[\texttt{MESA};][]{Paxton_2011,Paxton_2013,Paxton_2015,Paxton_2018,Paxton_2019,Jermyn_2023}, to model massive stars at solar metallicity $Z_\odot=0.0142$~\citep{Asplund_2009}. In order to focus on \gls{BH} progenitors, we simulate stars with initial masses between 20 and 300~$M_\odot$ with a sampling width of 5~$M_\odot$. This mass range was selected in order to investigate specifically the \gls{BH} progenitor parameter space. Table~\ref{tab:models} shows a summary of all the different model assumptions for stellar winds.

\paragraph{\textbf{Structure, mixing, rotation}} We adopt \cite{Ledoux_1947} for convective boundaries and mixing length $l= 1.82$ \citep{Choi_2016MIST}. We use exponential overshooting with $f_{\rm ov}=0.05$ \citep{Scott_2021} above every convective region. While a lower overshooting efficiency is typically required to match the observed population at lower masses, we adopt this value for all stars with $M_{\rm ZAMS} \geq 20~M_\odot$. This choice is supported by the work of \cite{Scott_2021}, which suggests that such higher values are consistent with the structural properties of more massive progenitors. This choice is also further motivated by the fact that the final \gls{BH} mass from the evolution of isolated rotating $Z_\odot$ stars becomes relatively insensitive at $f_{\rm ov}\geq0.04$ \citep{Romagnolo_2024}.
We also adopt a value of exponential undershooting of $f_{\rm under}=f_{\rm ov}/10$. 3D simulations show that the formation of semiconvective regions could be completely suppressed due to strong overshooting \citep{Blouin_2023}, and we therefore do not include semiconvection in our models \citep[for a detailed evolutionary parameter space investigation across different combinations of overshooting and semiconvection values, we refer to][]{Schootemeijer_2019}. To reduce superadiabaticity in regions near the Eddington limit, we adopt the \textit{use\_superad\_reduction} method \citep{Jermyn_2023}. We set the initial angular velocity to $0.4$ of the critical velocity. To model angular momentum and chemical mixing, we use the calibrations from \cite{Heger_2000}, with the addition of Tayler-Spruit dynamo \citep{Tayler_1973,Spruit_2002,Heger_2005}. We adopt the default \cite{Heger_2000} {\tt MESA} rotational mass loss enhancement.

\paragraph{\textbf{Core collapse}} We stop all our simulations at core-C depletion ($M_{\rm core, C}$/$M_{\rm core}<$\,10$^{\rm -2}$), since the time to core-collapse is negligible in terms of wind-driven mass loss \citep[e.g.][]{Hirschi_2025}\footnote{After core-carbon burning, the timescale before core-collapse is measured in years \citep{Woosley_2002}. Even overestimating wind-driven mass loss by orders of magnitude to e.g. $10^{-2}~M_\odot yr^{-1}$ would only lead to a total post-core-carbon burning mass loss in the order of 0.01~$M_\odot$ .}. In order to model nuclear reactions until this stage, we use the \textit{approx21\_mn57\_plus\_co56} nuclear network in {\tt MESA}. As a first-order approximation, we use an upper limit to the \gls{BH} mass and assume that all stars face direct \gls{BH} collapse, conserving the totality of their pre-core collapse mass minus 1\% being ejected as neutrinos\footnote{While we assume direct collapse, the neutrino mass loss can initiate a shockwave leading to the ejection of \gls{RSG} envelopes \citep{Nadezhin_1980,Lovegrove_2013}. This effect is more considerable for \glspl{RSG} than \glspl{WR} \citep{Fernandez_2018}, and could further reduce the final mass of supergiant remnants.} \citep{Belczynski_2020}. While this physical assumption is reasonable for compact He stars, it may considerably overestimate the \gls{BH} masses of lower-mass progenitors that retain a loosely bound H-rich envelope, as this envelope is highly susceptible to ejection. Furthermore, we explicitly ignore the effects of rotation on the collapse dynamics. For progenitors with retained envelopes, the assumption of direct collapse therefore represents a strict higher mass boundary for the compact object that can be formed. Finally, we do not include the effects of \gls{PPSN} and pair-instability supernova in our calculations.

\paragraph{\textbf{Stellar winds}}
In this work, we explore 14 different stellar wind models, each representing a commonly assumed wind combination from the literature and described in the following subsections. We detail the specifics below for each one (see Table~\ref{tab:models} for summary). 
To ensure consistency across all models, we standardize the scaling factor for the metallicity dependency by adopting a uniform solar metallicity of $Z_\odot$\,=\,0.0142 \citep{Asplund_2009} for all mass-loss recipes (see Sec.~$\S$3.3 of \AtlasI for more details).
Finally, as a reference, we assume as a first-order approximations that \gls{WR} stars are formed as soon as the progenitor star developes optically thick winds.





\subsubsection{Free-Electron Scattering transition (FESc) Models}
\label{subsubsec:FESc_family}

In our FESc models we use the classical Eddington factor based on free-electron scattering opacity, $\Gamma_{\rm e}$, as the evolutionary proxy to determine the onset of optically thick winds at $\Gamma_\text{e}\ge0.5$ \citep{Bestenlehner_2020}. This factor is a function of the surface hydrogen abundance ($X_\text{surf}$), luminosity and mass, and it is expressed as:

\begin{equation}\label{eq:logGammaEdd}
    \Gamma_{\rm e}\,=\,10^{\rm -4.813}(1+X_{\rm surf})(L/M)(L_\odot/M_\odot)
\end{equation}

For the main \textbf{FESc} scheme (`FESc' in Table~\ref{tab:models}) we calculate mass-loss rates in the following way:

\begin{itemize}
    \item \textit{Optically thin winds}: For $\log g\geq3.0$ we use \citet{GormazMatamala_2022b,GormazMatamala_2023}, hereafter GM23, while for $\log g<3.0$ we adopt \cite{Vink_2001}, hereafter V01. Finally, for non-\gls{WR} helium stars ($X_{\rm surf}\leq$\,0.4) we use \cite{Vink_2017}, hereby V17.
    
    \item \textit{Cool supergiant winds} ($T_{\rm eff}\leq$\,10~kK): \cite{deJager_1988}, hereby dJ88
    \item \textit{Optically thick winds}: for cool \glspl{WR} (10~kK\,$<T_{\rm eff}$\,$\leq$\,30~kK) we use V01. We then model H-rich \gls{WR} ($X_{\rm surf}>$\,10$^{\rm -7}$) winds with \citet{Bestenlehner_2020} with metallicity scaling from \cite{Brands_2022}, hereafter B20, and H-free \gls{WR} ($X_{\rm surf}\leq$\,10$^{\rm -7}$) winds with \citet{Sander_2020}, hereafter SV20. 
\end{itemize}


\paragraph{$\mathbf{FESc_\mathbf{V01}}$}

This is a submodel for the FESc model that maximizes thin winds mass loss by applying V01 to the whole thin winds parameter space. This variation has been adopted to highlight how the $\Gamma_{\rm e}$-based transition to thick winds is impacted by stronger mass-loss rates.

\paragraph{$\mathbf{FESc_\mathbf{RSG}}$}

In this submodel dJ88 is only used for \glspl{RSG}, while for \glspl{YSG} at 4~kK\,$<T_{\rm eff}<$\,10~kK, where mass-loss rates are nearly completely unconstrained (\citealt{Nieuwenhuijzen_deJager_1990}; \AtlasI), thin winds are applied. This scheme has been developed to study the role of \gls{YSG} winds and the choices of winds in this parameter space.

\paragraph{$\mathbf{FESc_\mathbf{noHpoor}}$}

In this submodel we apply no dedicated H-free \gls{WR} mass loss scheme in order to study the impact of artifically underestimating mass loss for such stars. When $X_{\rm surf}$ reaches levels below $10^{\rm -7}$, mass-loss rates still follow B20.

\paragraph{$\mathbf{FESc_{highrot}}$}

In this submodel we apply a stronger initial rotation to study enhanced rotational mixing and mass loss. In this submodel we apply $\Omega_{\rm init}/\Omega_{\rm crit}\,=\,0.6$ (roughly equivalent to the {\tt GENEC} models at $V_{\rm init}/V_{\rm crit}=0.4$; see \citealt{GormazMatamala_2025}).

\paragraph{$\mathbf{FESc_\mathbf{A24}}$}

The only difference with the original model is that cool supergiant winds are modeled with the implementation of the updated weak steady state  wind \gls{RSG} formulae from \cite{Antoniadis_2024}, hereafter A24, within the range $T_{\rm eff}\leq 4$~kK and $\log L <$\,5.8. Between 4~kK and 10~kK, instead, or at $\log L \geq$\,5.8, cool supergiant winds still follow dJ88. We 
do not include model variations for other cool supergiant wind schemes, since their extrapolation to luminosities past the \gls{HD} limit may lead to wide uncertainties and artifacts, as described in \AtlasI.

\subsubsection{Multiple-Scattering transition (MSc) Models}

Our \textbf{main MSc model} (`MSc' in Table~\ref{tab:models}) adopts the  wind efficiency parameter to determine the transition to optically thick winds:

\begin{equation}
    \eta \equiv \dot{M}\varv_{\infty}/(L/c)    
\end{equation}

Where $\varv_{\infty}$ is the terminal velocity and \textit{c} the speed of light.
The transition to the optically thick regime occurs when $\eta$ exceeds a specific threshold defined by $\eta_{\rm trans} = 0.75 \left(1 + \frac{\varv_{\rm esc}^2}{\varv_{\infty}^2}\right)^{-1}$ \citep[calibration from][]{Sabhahit_2023}, with $\varv_{\rm esc}$ the escape velocity. We refer to \AtlasI and our open-source models\footnote{\url{https://github.com/AmedeoRom/Stellar_Winds_Atlas}} for the detailed implementation of these parameters.

This model is otherwise identical to FESc, with the only exception being that if $\Gamma_{\rm e} < 0.4$ during the optically thick phase, we still apply V01 to avoid extrapolation artifacts \citep{GormazMatamala_2022}. These choices were made to be consistent with the FESc models and have a constrained comparison between the two transition conditions.


\paragraph{$\mathbf{MSc_\mathbf{RSG}}$}

Similary to the FESc$_{\rm RSG}$ submodel, we want to show the role of cool supergiant mass-loss rates when only applied to \gls{RSG}s. We therefore modify MSc with dJ88 only being applied for $T_{\rm eff}<$\,4~kK to get to a closer resemblance to the \cite{Vink_2024} models.

\paragraph{$\mathbf{MSc_\mathbf{V01}}$}

This submodel adopts V01 for all thin winds and \cite{Vink_2011}, hereby V11, for thick winds. For consistency with \cite{Vink_2024}, this submodel also adopts dJ88 only at $T_{\rm eff}$\,$\leq$\,4~kK, SV20 for H-free \gls{WR}, and has no switch to optically thin mass-loss rates for $T_{\rm eff}$\,$\leq$\,30~kK.

\subsubsection{$X_{\rm surf}$ Model}

With this model, we want to study the evolution of stars with the $X_{\rm surf}$ transition to thick winds. In this case, everything is the same as FESc, but models enter the thick winds regime only once $X_{\rm surf}<$\,0.4 .

\subsubsection{Dutch-ish Model}
\label{subsubsec:Dutch_family}

In this model we provide a mass-loss scheme that closely resembles one of the so-called Dutch wind models. The only major difference from the canonical Dutch winds is the implementation of GM23 for $\log g > 3$, while V01 is initiated only with lower surface gravity. We transition to thick winds at $X_{\rm surf}<$\,0.4, and mass-loss rates beyond this transition follow \cite{NugisLamers_2000}, hereby NL00, even for cool \gls{WR} stars. For H-free \gls{WR} stars, instead, we use the mass-loss rates from Eq.~21 of \cite{NugisLamers_2000}, with the calibrations from \cite{Eldridge_2006}, hereby EV06.

\paragraph{\textbf{Dutch}} This refers to the widely used ``Dutch winds'' scheme, with V01 for the whole optically thin regime. This is the model with the strongest mass loss \citep[see also][]{Pauli_2025}.

\subsubsection{KABS Model}
\label{subsubsec:KABS}


The \textit{Krticka, Antoniadis, Bestenlehner, Sander} (KABS) model was originally developed in \AtlasI, representing the weakest mass-loss rates for line-driven and \gls{RSG} winds in our sample. For the optically thin and cool \gls{WR} winds, we adopt \cite{Krticka_2024}, hereby K24. Additionally, we do not adopt only one transition condition. We apply B20 as soon as $\eta>\,\eta_{\rm trans}$ or $\Gamma_{\rm e}>\,0.5$, SV20 for H-free \glspl{WR}, and V17 for non-\gls{WR} He-stars. For cool supergiant winds, instead, we use dJ88 for 10~kK\,$>\,T_{\rm eff}\,>$\,4~kK or $T_{\rm eff}\,<$\,10~kK and $\log L\,>$\,5.8, and A24 for $T_{\rm eff}\,<$\,4~kK and $\log L\,<$\,5.8 .

\subsubsection{Warm\_$\Gamma$ Model}

This model is identical to the KABS model, but we adopt \cite{Pauli_2025}, hereby P25, for the optically thin, cool \gls{WR}, and non-\gls{WR} He star regime. This model was created to have a fully consistent formulation as a function of $\Gamma_{\rm e}$ for line-driven winds, without any bistability jump. Although P25 was also calibrated for optically thick winds from \gls{WR} star observations, the authors show no mass-loss rate increase within the optically thick regime, which they attribute to their limited sample of \gls{WR} binaries available for their fits. This motivates our choice to use P25 only for cool \gls{WR} winds.

\subsection{Population Synthesis}
\label{subsec:popsynth}

To test our evolutionary models for \glspl{VMS}, i.e. massive stars that develope optically thick winds already during \gls{MS}, we use the new Galactic population synthesis code {\tt StarEstate}\footnote{\url{https://github.com/AmedeoRom/StarEstate}} \citep{Romagnolo_2025_StarEstate}. The code can sample stars from {\tt MESA} simulations at given initial distributions of $M_{\rm ZAMS}$, \textit{Z}, lookback time and Galactic position. We use this code to generate synthetic Galactic populations of $M_{\rm ZAMS}$\,$\geq$\,50~$M_\odot$ stars from MSc, Dutch, MSc$_{\rm V01}$, FESc$_{\rm V01}$ (Section~\ref{subsec:thick_trans}), and Warm\_$\Gamma$. We do so by creating a Milky Way model at 5.2\,$\times$\,10$^{\rm 9}$~$M_\odot$, which roughly represents 10\% the total Galactic thin disk mass \citep{Licquia_2015}. Following \cite{Wagg_2022}, we then adopt an exponentially declining star formation history and the metallicity-radius-time relation from \cite{Frankel_2018}, with a truncated normal distribution of metallicities up to $Z$\,=\,0.03 . Across the whole metallicity distribution, we bin all the stars with initial \textit{Z} within 20\% from $Z_\odot$ into our Milky Way sample. We then draw the binary population assuming a \gls{VMS} binary fraction of 75\% (i.e. 6 out of 7 stars in binaries), the companion initial mass $M_{\rm ZAMS, 2}$ from a uniform distribution between 20~$M_\odot$ and the initial mass of the primary companion $M_{\rm ZAMS, 1}$, and the initial orbital period (\textit{P}, $\log (P/d) \in [0.5; 5.5]$) and eccentricity ($e \in [0.0; 0.9]$) from power law distributions, respectively, with exponents $\alpha_P = -0.55$ and $\alpha_e = -0.42$ \citep{Sana_2012,deMink_2015}. 

We then evaluate the occurrence of \gls{RLOF} across our models by integrating the binaries' orbital evolution over time. This integration accounts for wind-driven mass loss and tidal interactions \citep{Hut_1981} as implemented in \cite{Hurley_2002}, tracking the periastron distance and the Roche-lobe radius \citep{Eggleton_1983}\footnote{We highlight that \gls{RLOF} may be initiated before the optically thick stellar atmospheres expand past their Roche lobe, but this cannot be estimated for {\tt MESA} post-processing due to the lack of a self-consistent calculation for optically thin atmospheric radii \citep{Ritter_1988,KolbRitter_1990}}. Systems that successfully avoid \gls{RLOF} throughout their entire evolutionary history are classified as non-interacting. 

For all the single and effectively single \glspl{VMS} in our sample, we use as a comparison the observed population of OB and H-rich \gls{WR} stars in the Milky Way at $\log(L/L_\odot)\geq5.5$, from the catalogs of \cite{Martins_2008}, \citet{Hamann_2019}, and \citet{Pauli_2025}, hereby M08, H19, P25.

\section{Results}

\subsection{Black hole masses as a function of initial stellar mass}
\label{sec:results_BH}

\begin{figure*}[p]
\centering
\includegraphics[width=0.98\textwidth]{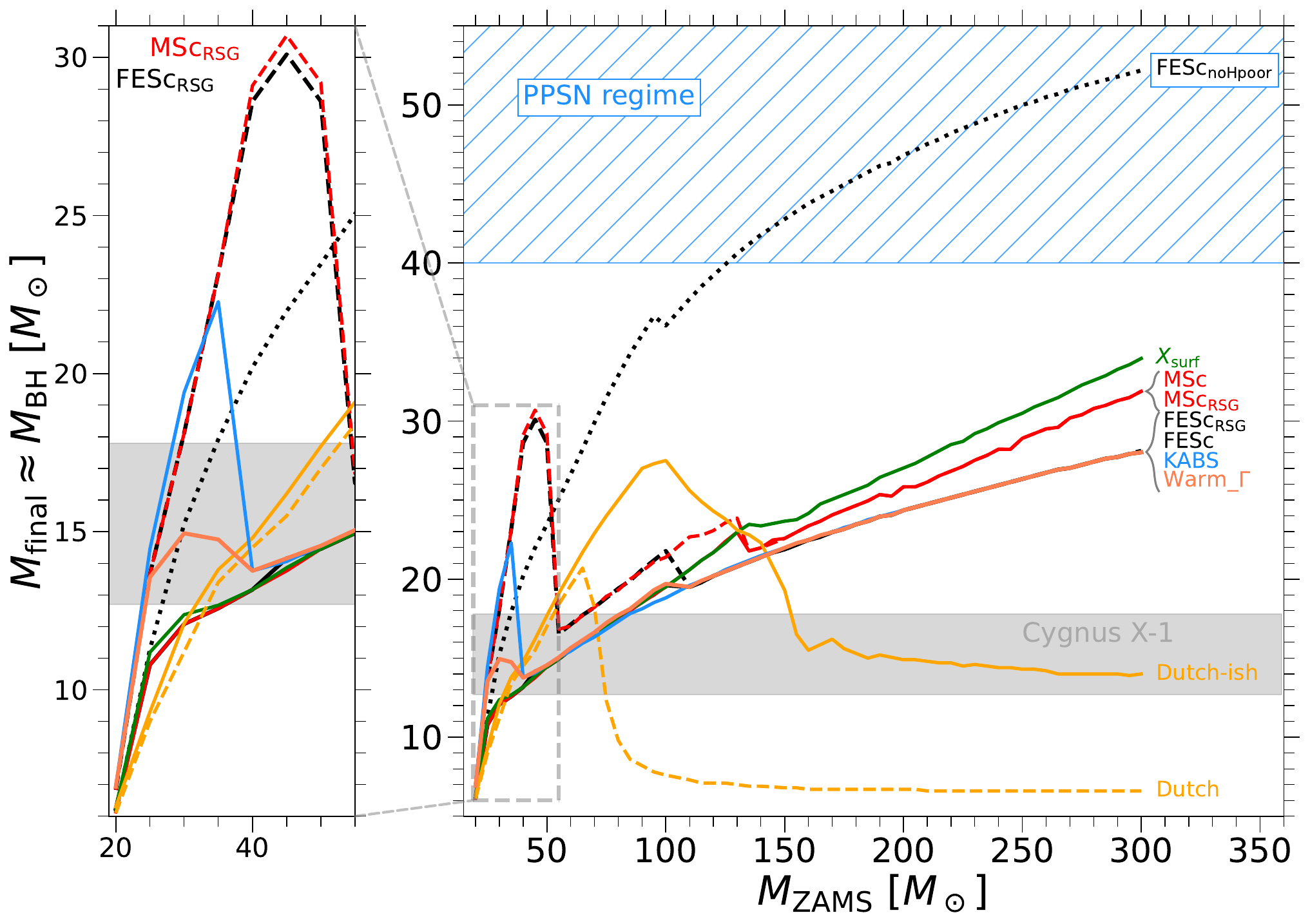}
\caption{$M_{\rm BH}$ as a function of $M_{\rm ZAMS}$ for our models. The mass range of Cygnus~X-1 \gls{BH} \citep{Ramachandran_2025} in grey. The left figure represents a zoomed-in plot within $M_{\rm ZAMS}\,=\,55$~$M_\odot$. The production of massive \gls{BH}s at $M_{\rm ZAMS}\,\lesssim\,50$~$M_\odot$ is only achievable with weak mass loss during the cool supergiant regime (FESc$_{\rm RSG}$, MSc$_{\rm RSG}$, KABS models). Only models that considerably underestimate H-free \gls{WR} mass-loss rates (FESc$_{\rm noHpoor}$) can enter the \gls{PPSN} regime at this metallicity.
    }
\label{fig:BHmass}
\end{figure*}

Figure~\ref{fig:BHmass} shows the \gls{BH} mass ($M_{\rm BH}$) as a function of $M_{\rm ZAMS}$, for each of the presented evolutionary models. We show that \gls{BH} masses are deeply susceptible to the adopted wind models, with negligible contributions between $\Omega_{\rm init}/\Omega_{\rm crit}=0.4$ and $\Omega_{\rm init}/\Omega_{\rm crit}=0.6$, since angular momentum is quickly removed due to strong mass loss. We will now discuss some of the most prevalent results.

\begingroup
\renewcommand{\arraystretch}{1.4}
\begin{table*}
\centering
\caption{Adopted models with respective initial conditions, divided among the adopted conditions for thick winds initiation \label{tab:models}}.
\begin{tabular}{l|cc|cc|cccc|c}
\hline
Model & log$g\geq3$ Thin & log$g<3$ & $T_{\rm eff}\leq$10~kK & $T_{\rm eff}\leq$4~kK & Thick Winds & Cool H-rich & H-rich & H-free & Comments\\
\hline
    \textbf{FESc} & GM23 & V01 & dJ88 & dJ88 & $\Gamma_{\rm e}\geq0.5$ & V01 & B20 & SV20 & --\\
    \textbf{FESc$_{\rm RSG}$} & GM23 & V01 & V01 & dJ88 & $\Gamma_{\rm e}\geq0.5$ & V01 & B20 & SV20 & --\\
    \textbf{FESc$_{\rm V01}$} & V01 & V01 & V01 & dJ88 & $\Gamma_{\rm e}\geq0.5$ & V01 & V11 & SV20 & Only in Section~\ref{subsec:thick_trans}\\
    \textbf{FESc$_{\rm highrot}$} & GM23 & V01 & dJ88 & dJ88 & $\Gamma_{\rm e}\geq0.5$ & V01 & B20 & SV20 & $\Omega_{\rm init}/\Omega_{\rm crit}$\,=\,0.6\\
    \textbf{FESc$_{\rm noHpoor}$} & GM23 & V01 & dJ88 & dJ88 & $\Gamma_{\rm e}\geq0.5$ & V01 & B20 & B20 & --\\
    \textbf{FESc$_{\rm A24}$} & GM23 & V01 & dJ88 & A24 & $\Gamma_{\rm e}\geq0.5$ & V01 & B20 & SV20 & --\\
    \tableline 
    \textbf{\color{red}MSc} & GM23 & V01 & dJ88 & dJ88 & $\eta\geq\eta_{\rm trans}$ & V01 & B20 & SV20 & --\\
    \textbf{\color{red}MSc$_{\rm RSG}$} & GM23 & V01 & V01 & dJ88 & $\eta\geq\eta_{\rm trans}$ & V01 & B20 & SV20 & --\\
    \textbf{\color{red}MSc$_{\rm V01}$} & V01 & V01 & V01 & dJ88 & $\eta\geq\eta_{\rm trans}$ & V11 & V11 & SV20 & Only in Section~\ref{subsec:thick_trans}\\
    \tableline
    \textbf{\color{PineGreen}$X_{\rm surf}$} & GM23 & V01 & dJ88 & dJ88 & $X_{\rm surf}<0.4$ & V01 & B20 & SV20 & --\\
    \textbf{\color{orange}Dutch-ish} & GM23 & V01 & dJ88 & dJ88 & $X_{\rm surf}<0.4$ & NL00 & NL00 & EV06 & --\\
    {\color{orange}\textbf{Dutch}$^1$} & V01 & V01 & dJ88 & dJ88 & $X_{\rm surf}<0.4$ & NL00 & NL00 & EV06 & --\\
    \tableline
    \textbf{\color{Cerulean}KABS}\tablenotemark{*}  & K24 & K24 & dJ88 & A24 & $\Gamma_{\rm e}\geq0.5$; $\eta\geq\eta_{\rm trans}$ & K24 & B20 & SV20 & --\\
    \textbf{\color{coral}Warm\_$\Gamma$}\tablenotemark{**} & P25 & P25 & dJ88 & A24 & $\Gamma_{\rm e}\geq0.5$; $\eta\geq\eta_{\rm trans}$ & P25 & B20 & SV20 & --\\
\hline
\end{tabular}

\parbox{\textwidth}{\footnotesize
    \textsuperscript{1}{Unlike the default {\tt MESA} Dutch formulation, the $Z_\odot$ scaling factor is set to 0.0142, rather than being hard-coded at 0.019.}\\
    \textsuperscript{*}{Fiducial model with bistability jump. Lowest optically thin $\dot{M}$.}\\
    \textsuperscript{**}{Fiducial model without bistability jump.}\\
    \textbf{References}: dJ88 -- \citet{deJager_1988}; NL00 -- \citet{NugisLamers_2000}; V01 -- \citet{Vink_2001}; EV06 -- \citet{Eldridge_2006}; V11 -- \citet{Vink_2011}; B20 -- \citet{Bestenlehner_2020}; SV20 -- \citet{Sander_2020}; GM23 -- \citet{GormazMatamala_2023}; A24 -- \citet{Antoniadis_2024};  K24 -- \citet{Krticka_2024}; P25 -- \citet{Pauli_2025}\\
}

\end{table*}

\endgroup

\subsubsection{The BH mass peaks from strong WR winds (Dutch models)} 

As shown in Figure~\ref{fig:BHmass}, the Dutch and Dutch-ish models are the only ones predicting a peak in the $M_{\rm BH}$-$M_{\rm ZAMS}$ relation, followed by a sharp decline at higher initial masses \citep[see also][]{Ekstrom_2012,Bavera_2023,Kruckow_2024}. While the peak's precise location in mass is considerably model-dependent, the qualitative shape of this distribution is preserved. For these specific models, the subsequent drop in $M_{\rm BH}$ is caused by an early transition to optically thick winds during \gls{MS}. For stars above a certain $M_{\rm ZAMS}$ threshold, the combined effects of mass loss and internal mixing reduce $X_{\rm surf}$ sufficiently to trigger this transition. Consequently, these stars spend a significant portion of their \gls{MS} lifetime under NL00, which is known to considerably overestimate mass-loss rates \citep{Pauli_2025}. Since the NL00 rates are proportional to the surface helium abundance \textit{Y}, mass loss accelerates as the star evolves through the \gls{MS}, stripping mass so efficiently that \glspl{VMS} form considerably less massive \glspl{BH}. This feature also explains the similar qualitative behavior displayed for \gls{BH} masses from \cite{Vink_2024}, since the authors adopt V11 for optically thick winds, which was shown in \AtlasI to lead to mass-loss rates that are comparable in strength or stronger than NL00 at high luminosity levels.

\subsubsection{The Cool Supergiant BH mass peaks}
\label{subsec:BH_peak_RSG}

Figure~\ref{fig:BHmass} also shows a zoomed-in plot of the \gls{BH} masses from single stars of $M_{\rm ZAMS}\in[20, 55]~M_\odot$.
The formation of a $\sim$\,30~$M_\odot$ \gls{BH} peak within this range is only limited to models that do not initiate cool supergiant winds for non-\gls{RSG} stars ($T_{\rm eff}\gtrsim4$~kK). Despite there being a noticeable dependency on the adopted optically thin and cool supergiant mass-loss rates (see the KABS and Warm\_$\Gamma$ models), the main driver for the appearance of such a peak is that $Z_\odot$ post-\gls{MS} stars spend a negligible amount of time during their rapidly expanding \gls{HG} phase (see also \citealt{Zapartas_2025}). The choice of cool supergiant winds becomes instead important for when the star will enter \gls{CHeB}. Only stars at $M_{\rm ZAMS}\,\lesssim\,30$~$M_\odot$ enter \gls{CHeB} as \gls{RSG}s \citep[see also][]{Ekstrom_2025}, while for higher masses, optically thin winds are strong enough to limit the star in its \gls{YSG} phase, but not enough to move the star towards the \gls{WR} branch and form a less massive \gls{BH}s. 

Furthermore, our results show that the choice of the transition condition to optically thick winds is nearly irrelevant for the formation of \gls{BH}s within this mass range, in contrast to the claims of \cite{Vink_2024}. This is due to the fact that all the transitions occur nearly simultaneously due to the depletion of the H-rich layer during the cool supergiant phase (see for more details Section~\ref{subsec:thick_trans}).

\subsubsection{The cool supergiant BH mass peak and its sensitivity to wind prescriptions}
\label{subsubsec:40_var}

The existence of a \gls{BH} mass peak around \,30~$M_\odot$ at $M_{\rm ZAMS}$\,$\sim$\,40~$M_\odot$ is highly sensitive to cool supergiant winds. The primary factor controlling this feature is the effective temperature threshold ($T_{\rm eff, trans}$) at which cool supergiant winds are initiated. The peak is suppressed for strong \gls{YSG} winds, as the star's entire hydrogen envelope gets stripped. This process forces the progenitor to evolve into a \gls{WR} star and ultimately forming a lower-mass \gls{BH}.

\begin{figure}[!ht]
\centering
\includegraphics[width=0.485\textwidth]{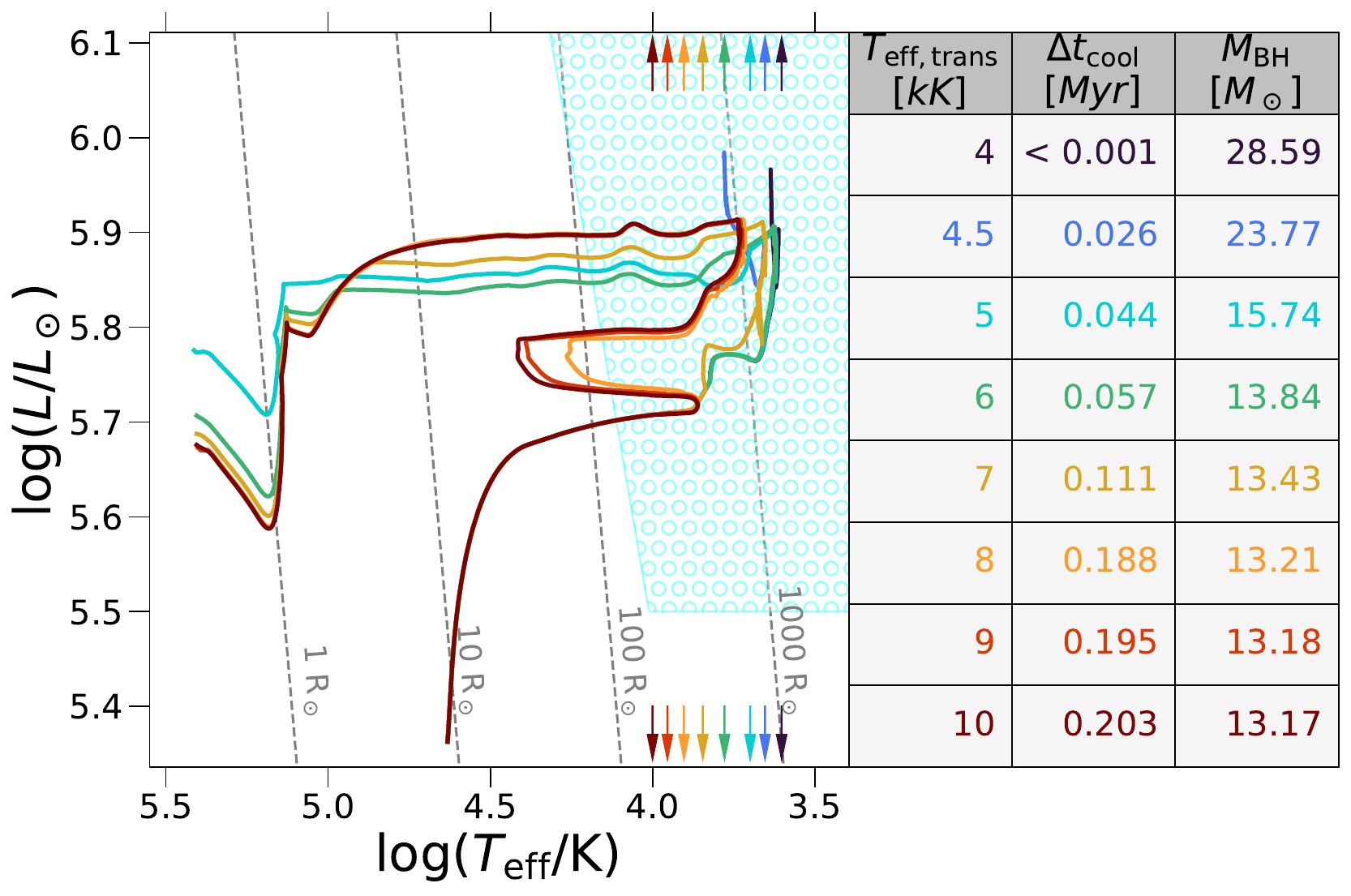}
\caption{Left \gls{HR} diagram of a 40~$M_\odot$ star at $Z_\odot$ with the FESc model. Each color shows the evolution of the star with a different $T_{\rm eff, trans}$ to cool supergiant winds (see both the table and the arrows). The cyan-shaded region represents the \gls{HR} region beyond the \gls{HD} limit where stars are \glspl{LBV}. Right: table listing, with the same colors of the evolutionary tracks, the transition temperature to dJ88, the cumulative cool supergiant wind time, and final \gls{BH}. At $T_{\rm eff, trans}\,>$\,5~kK the star spends enough time in its cool supergiant phase to eject the totality of its envelope, become a \gls{WR} star, and collapse into a lower-mass \gls{BH}.}
\label{fig:40_variable_HR}
\end{figure}

Figure~\ref{fig:40_variable_HR} shows the evolution of a FESc 40~$M_\odot$ star under different $T_{\rm eff, trans}$  values. If cool supergiant winds are restricted to \gls{RSG}s ($T_{\rm eff, trans}$\,$\sim$\,4~kK), the time spent in this phase ($\Delta t_{\rm cool}$) is negligible, and the star collapses into a massive \gls{BH} of 28.6~$M_\odot$. Increasing the threshold by just 0.5~kK to 4.5~kK initiates a brief cool supergiant phase (0.026~Myr) that is sufficient to lower $M_{\rm BH}$ by nearly 5~$M_\odot$. Once $T_{\rm eff, trans}$ is set to 5~kK or higher, the mass loss is strong enough to completely remove the envelope, forming a \gls{WR} star and producing a considerably lower-mass \gls{BH} ($\sim$15.7~$M_\odot$). The final \gls{BH} mass then plateaus at $T_{\rm eff, trans}$\,$>$\,6~kK, changing by only a few tenths of $M_\odot$. 

This shows that the most critical factor is whether the star enters the \gls{WR} stage: once this transition occurs, thick winds dominate the subsequent mass loss, making the final outcome less sensitive to the preceding mass-loss history.
Even if \gls{YSG}s were assumed to have either thin or weaker cool supergiant winds, an alternative mechanism may still suppress the high-mass \gls{BH} peak. In our models, all evolutionary tracks for these stars cross the \gls{HD} limit, where they are expected to enter a \gls{LBV} phase with enhanced mass loss. The existence of the \gls{HD} limit itself implies that a more powerful mass loss mechanism might be active to prevent stars from evolving past this boundary \citep{Gilkis_2021,Pauli_2026}. If strong \gls{LBV} eruptions are responsible for enforcing the \gls{HD} limit, they would strip the stellar envelope and form a \gls{WR} star, which leads to the formation of a lower-mass \gls{BH}.

\subsubsection{The pair-instability mass gap and H-free WR winds}

Figure~\ref{fig:BHmass} shows that the FESc$_{\rm noHpoor}$ model, which was developed to artificially underestimate mass loss for H-free \gls{WR} stars, leading \glspl{VMS} to enter the \gls{PPSN} regime due to high CO core masses. However, none of the fiducial models get close to the onset of \gls{PPSN} due to higher H-free \gls{WR} mass-loss rates. This leads to the conclusion that if a \gls{BH} within the \gls{PPSN} regime was found in a high metallicity environment, this may rather be an indicator of past mass transfer or merger events within a binary star system, which may have considerably less massive cores compared to single-star evolution \citep{diCarlo_2019,diCarlo_2020,Schneider_2024}, therefore potentially overcoming the \gls{PPSN} limit \citep{Renzo_2020}.

\subsubsection{Comparison with observed BH masses: the case of~Cygnus X-1}
\label{subsec:cygX1}

Figure~\ref{fig:BHmass} shows the maximum \gls{BH} mass for different models, from observations the max \gls{BH} at solar metallicity is Cygnus~X-1, which is currently the observed highest-mass \gls{BH} with a $Z_\odot$ companion. Both the Cygnus~X-1 companion and \gls{BH} masses have undergone multiple revisions over time, with the latest estimates from \cite{Ramachandran_2025} placing the \gls{BH} mass at 1-$\sigma$ confidence interval between 12.7 and 17.8~$M_\odot$. This $M_{\rm BH}$ range is shown in Figure~\ref{fig:BHmass}.

Cygnus~X-1~BH aligns well with predictions of many models for $M_{\rm ZAMS}$\,$\lesssim$\,60~$M_\odot$. Given the conclusion of \cite{Ramachandran_2025} that the system has not yet experienced a \gls{RLOF} event prior to its current x-ray binary stage, we infer that the \gls{BH} originated from the nearly-isolated evolution of its progenitor. While the Dutch-ish model could also produce such a \gls{BH} at $M_{\rm ZAMS}$\,$\geq$\,150~$M_\odot$, we do not consider it our best guess for \gls{VMS} winds, since most of the implemented mass-loss rate formulae were observationally shown to considerably overestimate the strength of winds \citep[e.g.][]{Bouret_2005,Surlan_2012,Surlan_2013,Beasor_2023,Antoniadis_2024,Decin_2024,Pauli_2025}.

On the other hand, under the assumption of a past \gls{RLOF} phase, the \gls{BH} progenitor (donor) would have shed envelope mass, likely leading to a lower-mass \gls{BH}. In this scenario, the mass range for Cygnus~X-1 would serve as a lower-mass boundary. Nevertheless, a past \gls{RLOF} implies that \gls{VMS}s, which typically produce more massive \gls{BH}s in isolation according to most of our models, could be the progenitors of the Cygnus~X-1 \gls{BH}. This scenario presents unique challenges. As discussed in subsequent sections, many \gls{VMS}s in our models do not inflate or expand beyond 200~$R_\odot$, limiting the probability of a \gls{VMS} initiating \gls{RLOF} within Cygnus~X-1. Furthermore, even if a \gls{VMS} were to initiate \gls{RLOF}, the mass ratio between the two stars would likely lead to dynamical instability and the onset of a common envelope phase. Given that the \gls{VMS} donor would have been in its \gls{MS} phase (later stages evolve at much smaller radii), this would inevitably result in a stellar merger \citep{Belczynski_2008}, therefore precluding its evolution into an X-ray binary.

\subsection{Transition to optically thick winds and the Humphreys-Davidson limit}
       \label{subsec:thick_trans}

The transition to optically thick winds is deeply dependent on the star's mass-loss history. To investigate the model-dependency of this transition, we define the following parameters: $t_{\rm FESc}$, $t_{\rm FESc_{e, V01}}$,  $t_{\rm MSc}$, $t_{\rm MSc_{ V01}}$, $t_{X{\rm surf}}$, $t_\text{Dutch}$, $t_\text{KABS}$, and $t_\text{Warm}$, representing the stellar ages at which stars their respective models develop optically thick winds. For $t_{\rm MSc_{ V01}}$ we adopt both our standard metallicity scaling factor ($Z_\odot$\,=\,0.0142) and a variation at $Z_\odot$\,=\,0.019 to be more aligned with the original V01 formulation.

\begin{figure}[!ht]
\centering
\includegraphics[width=0.475\textwidth]{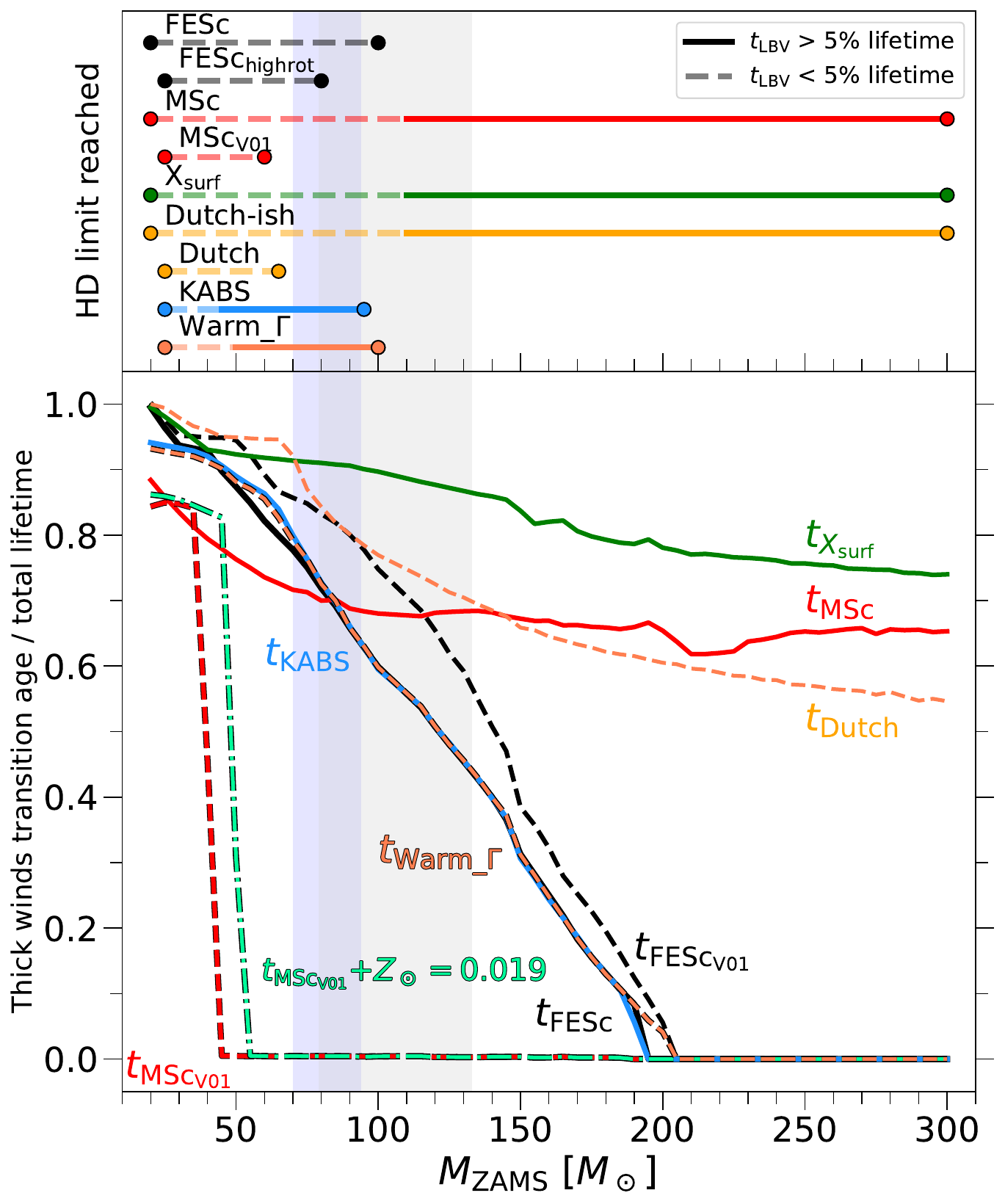}
\caption{Top: $M_{\rm ZAMS}$ ranges at which selected models do not respect the \gls{HD} limit. The full (dashed) lines indicate stars that stay \gls{LBV} for more (less) than 5\% of their total lifetime. Bottom: Ages at which stars transition to optically thick winds as a fraction of their total lifetime. The gray and blue areas show respectively the mass range for Cyg~OB2~\#12  \citep{Clark_2012} and WR~102a \citep{Lohr_2018}. The $\Gamma_{\rm e}\,\geq$\,0.5 transition to thick winds is nearly model-independent, while the $\eta\,\geq\,\eta_{\rm trans}$ transition strongly depends on the mass-loss history. Models with an early transition to thick winds or strong thin winds tend to better respect the \gls{HD} limit.}
\label{fig:thick_trans}
\end{figure}

Figure~\ref{fig:thick_trans} shows these transition ages, normalized by the stellar lifetime, as a function of $M_{\rm ZAMS}$, as well as the $M_{\rm ZAMS}$ range at which selected models present a transition beyond the \gls{HD} limit ($\log L$\,$\geq$\,5.5 and 10$^{\rm -5}\times R \times \sqrt{L}$\,$\geq$\,1), where we distinguish on whether the stars remain for an amount of time in their \gls{LBV} phase ($t_{\rm LBV}$), which we stress remains unmodeled in our analysis, that is more or less than 5\% of their total lifetime. 

\subsubsection{When do massive stars develop optically thick winds?}


At $M_{\rm ZAMS}\gtrsim$\,60~$M_\odot$, luminosity and mass are sufficient to drive $\Gamma_{\rm e}$ above the transition threshold during \gls{MS}, with the timing that is weakly dependent on the mass-loss history. This is evident from the fact that all the models that include a $\Gamma_{\rm e}$-driven condition to thick winds tend to enter the thick winds phase at roughly the same point in time, regardless of the strength of the thin and cool supergiant winds. On the other hand, the transition in the $X_{\rm surf}$, Dutch, and MSc models consistently occurs late in stellar evolution. In the MSc model, this is attributed to the low GM23 mass-loss rates, which hinder the increase of $\eta$ to the threshold for optically thick winds, as already hypothesized in \cite{GormazMatamala_2025}, considerably overproducing OB-type \glspl{VMS}.
In contrast, $t_\mathrm{MSc_{\rm V01}}$ shows a significantly earlier transition to the optically thick phase. The strong V01 thin winds lead stars with $M_{\rm ZAMS}\gtrsim$\,45~$M_\odot$ to enter the WNh phase near \gls{ZAMS}.
This comes in opposition to \cite{Vink_2024}, where it was claimed that the $\Gamma_{\rm e}$\,$\geq$\,0.5 transition (FESc models) would precede the $\eta$ transition for massive stars. $t_\mathrm{MSc_{\rm V01}}$ with $Z_\odot$\,=\,0.019, also predicts a near-\gls{ZAMS} transition, but at $M_{\rm ZAMS}\gtrsim$\,65~$M_\odot$ due to weaker mass loss (we refer to \AtlasI for more details on the role of the $Z_\odot$ calibrations).

This timing is further complicated by the fact that high mass-loss rates can actively delay the $\Gamma_{\rm e}$ transition if luminosity decreases proportionally faster than mass and $L/M$ decreases. This contrasts with the current theory of stellar winds. Higher mass-loss rates are supposed to lead to denser and thicker winds, and one would expect to observe \gls{WR}-type emission lines in the spectra, indicating the object is above the transition. However, we stress that using V01 and $Z_\odot$\,=\,0.0142 for the whole optically thin phase is already on the higher end for mass-loss rates, and the difference between $t_{\rm FESc_{V01}}$ and $t_{\rm FESc}$ is between 2\% and 15\%. This difference is far from negligible, but for $M_{\rm ZAMS}\gtrsim$\,50~$M_\odot$ stars, it is far lower than the one between $t_{\rm MSc}$ and $t_{\rm MSc_{ V01}}$, which can be roughly 5 to 37 times higher.


\subsubsection{Wind prescription, rotation, and the Humphreys-Davidson limit}


Following Figure~\ref{fig:thick_trans}, we find that whether a massive star evolves beyond the \gls{HD} limit is directly correlated with the timing of its transition to the optically thick phase, as also stated in \cite{Boco_2025}. The choice of wind prescription and transition criterion dictates this timing.

First, models that feature a delayed onset of strong mass loss allow stars to expand significantly after \gls{MS} and violate the \gls{HD} limit. This occurs in models with weak thin winds, especially when combined with the $\eta$ and $X_{\rm surf}$ conditions (MSc, $X_{\rm surf}$, and Dutch-ish models). By failing to quickly strip the H-rich envelope, these models predict stars crossing the \gls{HD} limit, with weaker thin winds usually leading to longer \gls{LBV} phases.

In contrast, models that trigger an early transition to thick winds keep stars compact, preventing them from crossing the \gls{HD} limit \footnote{This early transition can introduce a different conflict with observations. Initiating optically thick winds too soon prevents the evolutionary tracks from reproducing the observed population of cool supergiants.}. Models using the $\Gamma_{\rm e}$ transition respect instead the \gls{HD} limit for most \glspl{VMS} nearly regardless of their mass-loss history, with higher initial rotation (FESc$_{\rm highrot}$) further reducing the production of cool supergiants \citep{Gilkis_2021}.

Similarly, also the Dutch and $M_{\rm V01}$ models limit the formation of \gls{LBV}s due to the application of some of the strongest optically thin and thick winds in the literature \AtlasIp.

\subsection{Comparison with observed stellar populations}
\label{subsubsec:GalPop}

\begin{figure*}[!ht]
\centering
\includegraphics[width=0.9\textwidth]{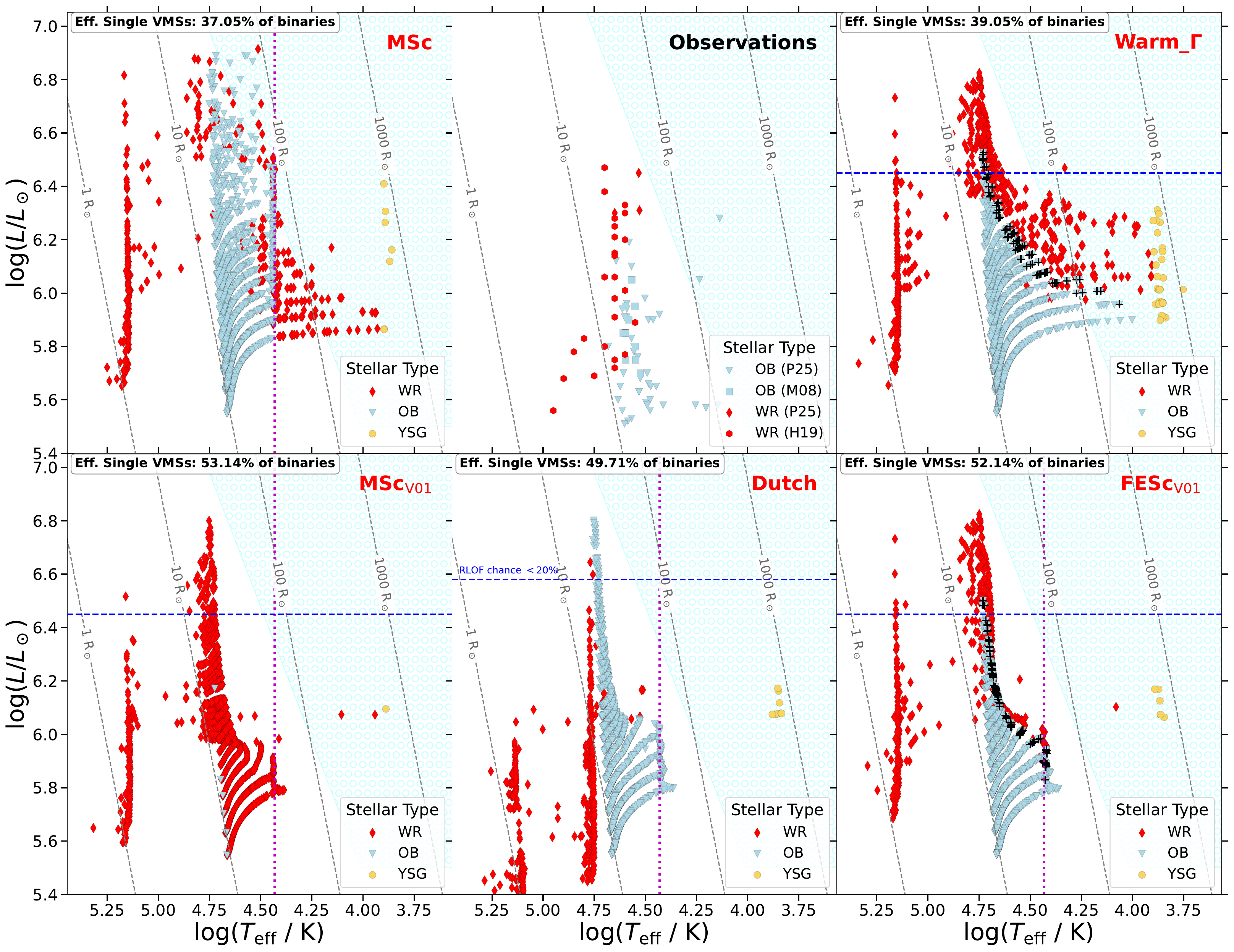}
\caption{Comparison between empirical Galactic observations and our synthetic populations of $M_{\rm ZAMS} \geq 50\,\Msun$ \glspl{VMS} scaled to 10\% of the Galactic thin disk mass. The synthetic datasets account for both single stars and binary components that successfully avoid mass transfer throughout their evolutionary history. The cyan-shaded region represents the \gls{LBV} regime, while the purple dotted vertical line indicates the $\log T_{\rm eff}$ threshold for the V01 bistability jump. Crossed symbols denote OB stars in the $\Gamma_{\rm e}$-based models where $\Gamma_{\rm e} \in [0.47; 0.5]$. Where applicable, blue horizontal dashed lines mark the critical luminosity threshold above which fewer than 20\% of \glspl{VMS} initiate \gls{RLOF} due to the limited \gls{VMS} expansion. The highlighted text cell in the top-left corner of each panel shows the percentage of binary systems whose components evolve in effective isolation. The MSc framework overpredicts the abundance of superluminous OB stars due to its combined weak thin winds and $\eta$-driven thick wind transition, whereas MSc$\mathrm{V01}$ underpredicts this subpopulation due to efficient thin winds. The $\Gamma_{\rm e}$-based models do not form OB stars at $\log (L/L_\odot) \gtrsim 6.5$, showing closer alignment with empirical constraints.
}
\label{fig:popsynt_comp}
\end{figure*}

The different model predictions for the timing for thick winds transition in Figure~\ref{fig:thick_trans} can be tested observationally, as we show in Figure ~\ref{fig:popsynt_comp}. The fraction of a star's life spent as an OB versus a \gls{WR} star \footnote{We stress that optically thick winds can also happen of OB stars, and that what models define as \gls{WR} does not necessarily represent an actual \gls{WR} star from spectroscopic observations.} should be reflected in the observed \gls{WR}/OB ratio at a given mass range. A model predicting a late transition implies a low \gls{WR}/OB ratio, whereas an early transition predicts a high one. Furthermore, if a model shows that beyond a certain $M_{\rm ZAMS}$ \gls{VMS}s are born as WNh stars, the observation of an OB star of a higher mass might weaken the validity of the model, despite not fully ruling it out, since other factors such as binary interactions add uncertainty in the conclusions. Additionally, it must be however highlighted that due to low-number statistics and the limited lifetime of such \glspl{VMS}, none of these models may be uniquely observationally verified by this methodology alone, and a cross-comparison with other observational benchmarks such as the \gls{BH} mass distribution in astrometric binaries, or comparing the theoretical and observational mass-loss rates for stars transitioning from the OB to the WN phase \citep[e.g.][for V01]{Vink_2012} is required.

As shown in Figure~\ref{fig:popsynt_comp}, the existence of OB stars like Cyg~OB2~\#12 with an mass of 110~$M_\odot$, is in tension with models that predict \gls{VMS}s becoming WNh stars near \gls{ZAMS} under that mass. Specifically, our models that combine V01 for all thin winds with the $\eta$ transition do not predict OB stars above  $\sim$45-65~$M_\odot$ at $Z_\odot$. Furthermore, despite Cyg~OB2~\#12 may host hidden stellar companions \citep{Rauw_2005,Albacete_Colombo_2007,Kiminki_2009} that may lead to an overestimate of the spectroscopic mass, no compelling observational evidence has yet been found for this scenario \citep{Clark_2012}. This suggests that models allowing for an extended OB phase at $M_{\rm ZAMS}$\,$\lesssim$\,150~$M_\odot$ are more consistent with \gls{VMS} observations.

In Figure~\ref{fig:popsynt_comp}, we compare our synthetic populations for $M_{\rm ZAMS} \geq 50\,\Msun$ against empirical Galactic observations of luminous OB and \gls{WR} stars. These massive single and binary systems account for $0.039\%$ of the integrated primary and secondary mass distribution.
When scaled to our specific Milky Way model, which isolates a subset corresponding to $10\%$ of the total Galactic thin disk mass, these targeted \glspl{VMS} comprise roughly $0.0039\%$ of the comprehensive Galactic stellar population. Crucially, the \glspl{VMS} displayed here are restricted exclusively to single and effectively single stars that escape binary interaction. Within our binary grids, these effectively isolated \glspl{VMS} represent a model-dependent fraction ranging from $\sim 27\%$ to $53\%$ of the total systems.

\paragraph{\textbf{Weak thin winds + $\eta$ transition: WR tension}} The MSc model predicts the most massive \glspl{VMS} to be beyond the \gls{HD} limit for long evolutionary times (Figure~\ref{fig:thick_trans}). Due to the use of both weak thin winds and $\eta$ transition, this model produces superluminous OB stars within a luminosity regime that has been so far observed to only host \gls{WR} objects in the Milky Way. This comes with the exception of the \gls{LBV} candidate $\eta$~Car~A, at $\log L\sim$\,6.4-6.8 \citep{Morris_2017}. However, evidence suggests that rather than being born as a single stellar object \citep{Kashi_2010}, $\eta$~Car~A is the result of a merger event between two \glspl{VMS} \citep{PortegiesZwart_2016}, making therefore its existence not a reliable factor in support of the MSc model.

\paragraph{\textbf{Strong thin winds + $\eta$ transition: OB tension}}
Contrary to the MSc model, adopting in MSc$_{\rm V01}$ stronger (V01) mass-loss rates leads to a population of \glspl{VMS} almost completely composed of \gls{WR} stars. This means that this specific model leads to an under-production of the number of OB stars. This suggests that a potential recalibration of the $\eta$ procedure might be needed for this prescription to fully represent the current observational constraints. It must be however highlighted that this result does not universally prove that the combination of V01 and $\eta$ transition is invalid. With different initial conditions for $\eta$, $Z_\odot$ scaling, and internal mixing, the models of \cite{Vink_2024} lead non-rotating solar metallicity \glspl{VMS} to evolve into WNh stars at \gls{ZAMS} only at $M_{\rm ZAMS}\gtrsim100~M_\odot$.

\paragraph{\textbf{Dutch models}} The Dutch model, like any other using V01 for the totality of the thin winds phase, nearly-completely avoids the forbidden region beyond the \gls{HD} limit, with a major bottleneck for the formation of cool supergiants being the bistability jump, beyond which nearly no star is shown in our population. However, despite the strong thin winds, the $X_{\rm surf}$ condition to thick winds limits a timely transition to the \gls{WR} stage, considerably overproducing the number of superluminous OB stars, as also highlighted in \cite{Pauli_2025}.

\paragraph{\textbf{$\Gamma_{\rm e}$ transition}} As we showed in Section~\ref{subsec:thick_trans}, the $\Gamma_{\rm e}$ transition to thick winds is weakly dependent on \gls{VMS} mass-loss history. This is further evident because the luminosity threshold at which FESc$_{\rm V01}$ and Warm\_$\Gamma$ beyond which no OB star is formed is nearly the same, in spite of model differences. The most luminous OB stars in these samples are predicted to be at $\log L\sim$\,6.5, while the most luminous OB star in the observational sample is at $\log L\sim$\,6.3 . Considering:

\begin{itemize}
    \item A typical error of $\sim$20\% in the luminosity estimates \citep{Markova_2018}
    \item That the highest-\textit{L} Warm\_$\Gamma$ OB stars are born at $\Gamma_{\rm e}$\,$\in$\,[0.47; 0.5], which is well within the uncertainty range for $\Gamma_{\rm e,trans}$ \citep{GormazMatamala_2025}
\end{itemize}

We consider the Warm\_$\Gamma$ model a potential match with the limited sample of the Galactic \gls{WR}/OB population. Furthermore, FESc$_{\rm V01}$ not only shows similar results in terms of Galactic \gls{WR}/OB populations, but also better respects the \gls{HD} limit due to the bottleneck of the V01 bistability jump. However, it must be stressed that the existence, strength, and \gls{HR} position of the bistability jump is under debate \citep{Bjorklund_2023,BerniniPeron_2024,deBurgos_2024,Verhamme_2024,Alkousa_2025,Krticka_2025,Pauli_2025,Verhamme_2026}, and this might not represent a valid venue to explain, even partially, the \gls{HD} limit, even considering that \gls{LBV} eruptions are potentially the strongest driver for the existence of such an observational threshold \citep[see e.g.][]{Cheng_2024, Pauli_2026}.

\subsection{Detailed evolutionary diagnostics}

Model uncertainties and their resulting variability must be understood beyond mere population-level trends, and it is necessary to examine the evolution of specific stars in detail within a common evolutionary framework. 
We provide here a detailed analysis of the mass and \gls{HR} evolution for a selected list of initial masses. We also use as a comparison the same observational catalogs that were shown in Section~\ref{subsubsec:GalPop}. We choose to study in detail the following stars:

\begin{itemize}
    \item 20~$M_\odot$: Lowest-mass star in our sample, and an exemplary case of a star evolving to the \gls{RSG} phase.
    \item 40~$M_\odot$: Lowest-mass star in our sample that does not enter the \gls{RSG} phase and roughly the $M_{\rm ZAMS}$ position at which an early \gls{BH} mass peak presents in Figure~\ref{fig:BHmass}.
    \item 75~$M_\odot$: A star that nearly in all models develops optically thick winds during \gls{MS} (Section~\ref{subsec:thick_trans}). Additionally, this roughly represents the most massive star for which rotational mixing still plays an important role. This is also roughly the $M_{\rm ZAMS}$ at which Dutch-like winds \citep[e.g.][]{Ekstrom_2012,Bavera_2023} predict the most massive \gls{BH}s to form at $Z_\odot$.
    \item 300~$M_\odot$: Highest-mass star in our sample. Depending on the assumptions on winds, such a star could already be born as a WNh with strong thick winds, or experience an initial OB phase prior to that point.
\end{itemize}

\subsubsection{Low-mass BH progenitors: the 20~$M_\odot$ case}

\begin{figure}[!h]
\centering
\includegraphics[width=0.42\textwidth]{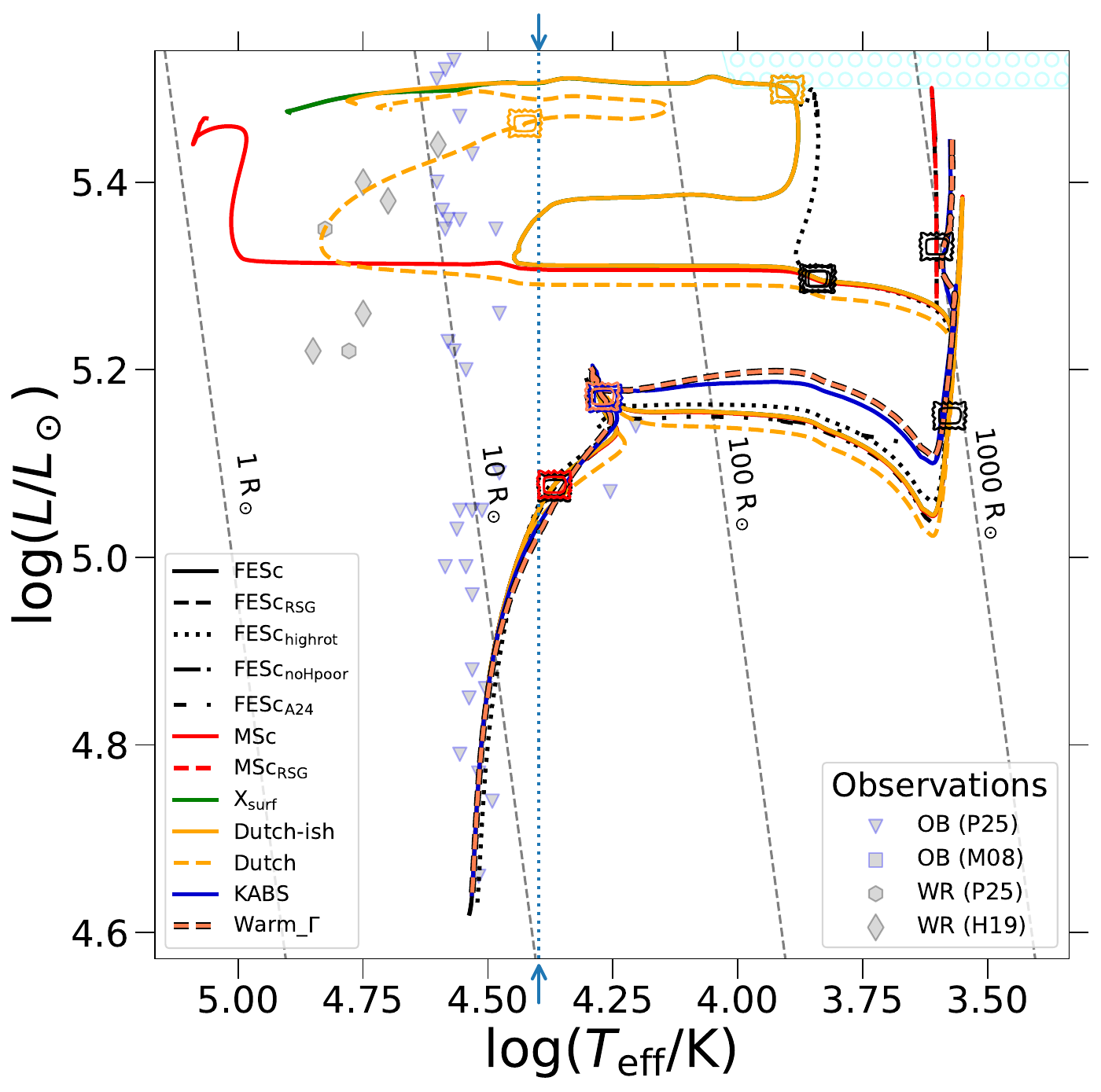}
\caption{HR diagram for $M_{\rm ZAMS}$\,=\,20~$M_\odot$ stars at $Z_\odot$ until the end of core-C burning. The arrows and the dotted vertical line indicate the $T_{\rm eff}$ for the V01 bistability jump. The ravioli scatter points instead indicate the point at which stellar winds become optically thick. The light blue region represents the \gls{LBV} area past the \gls{HD} limit.}
\label{fig:20_HR}
\end{figure}

\begin{figure}[!h]
\centering
\includegraphics[width=0.465\textwidth]{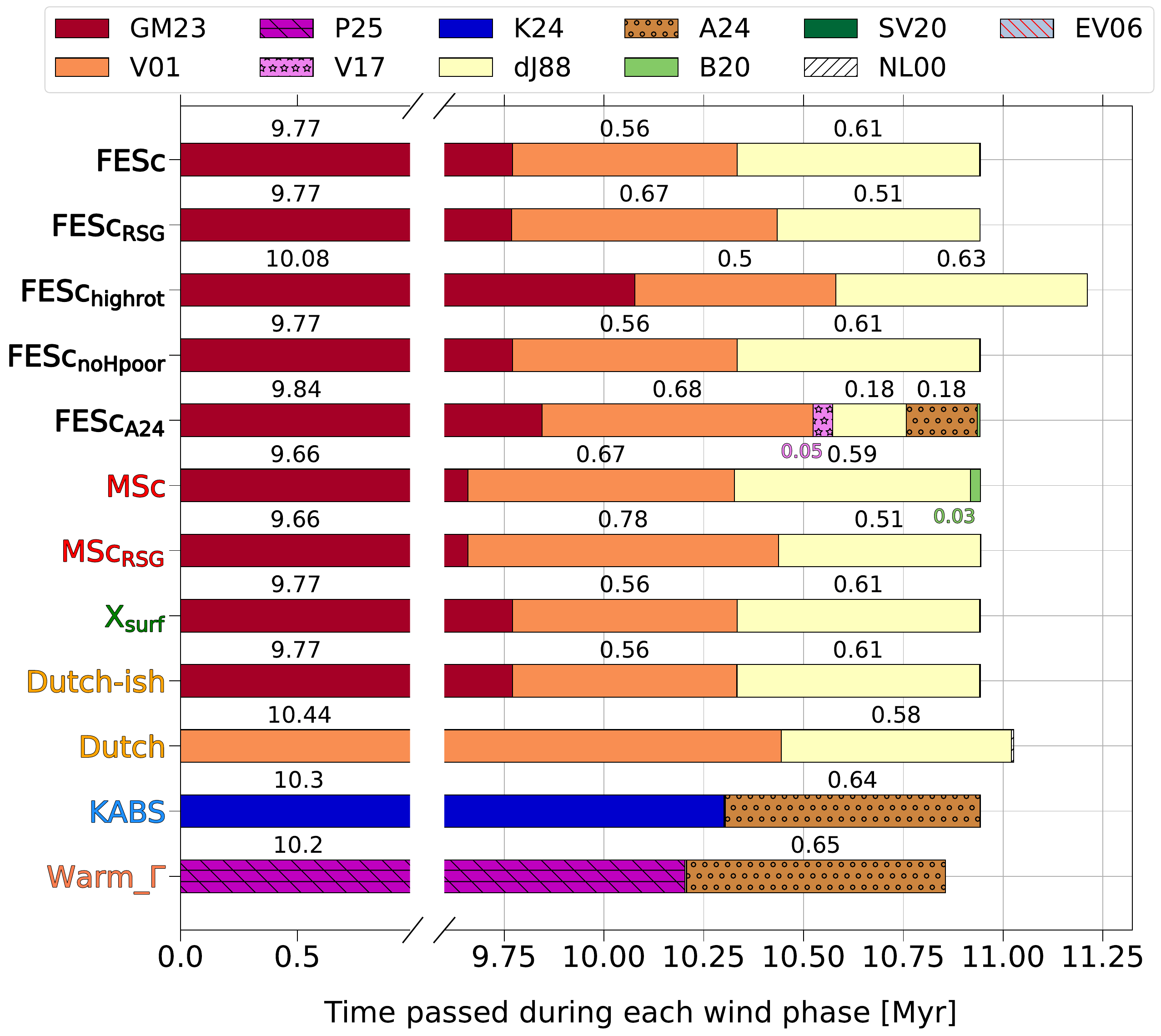}\\
\includegraphics[width=0.458\textwidth]{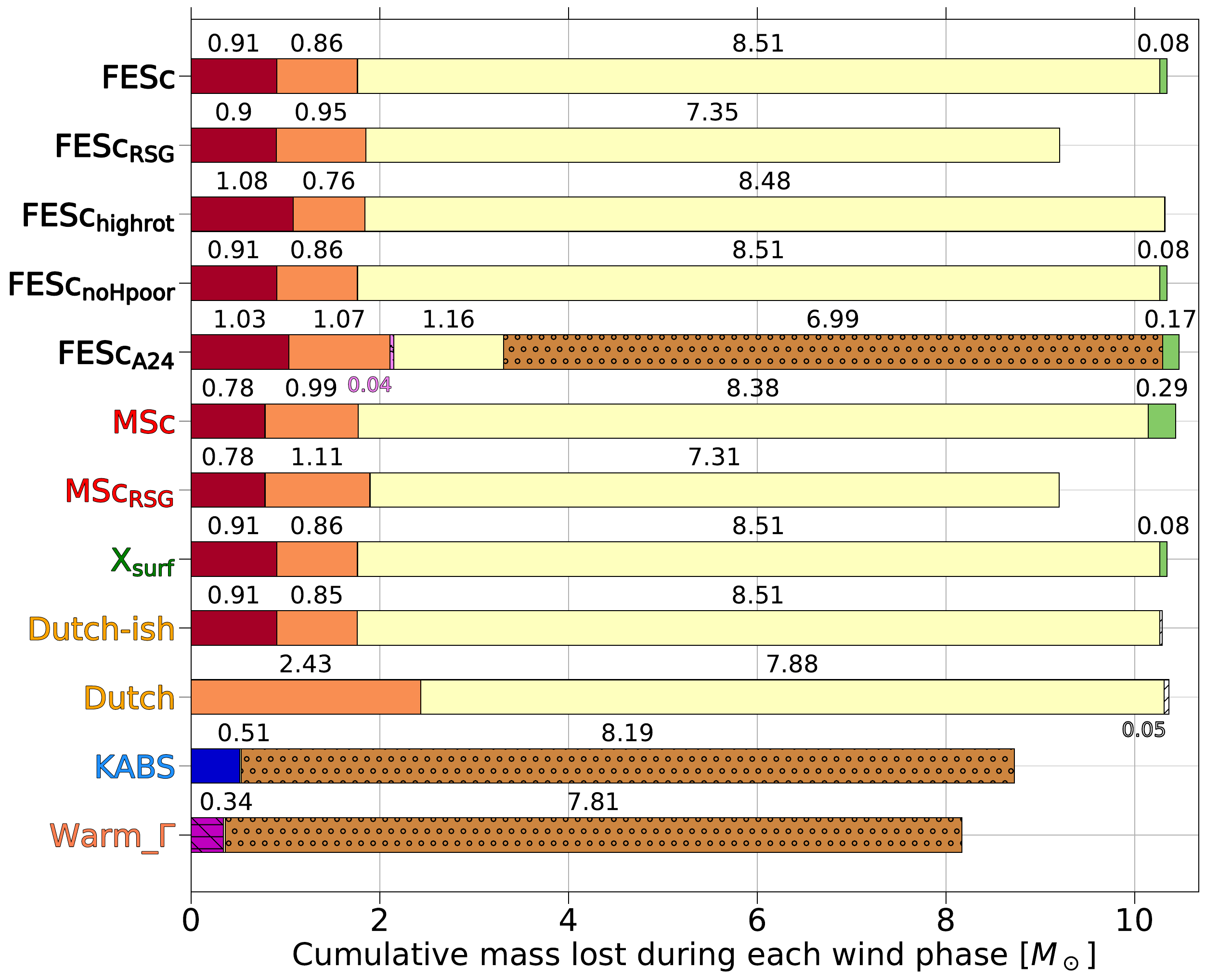}
\caption{Cumulative time and mass loss from each wind scheme for a $M_{\rm ZAMS}$\,=\,20~$M_\odot$ star at $Z_\odot$, according to each proposed evolutionary model. On the top plot, we cut the cumulative time plot to enhance the visualization of the whole evolution. Despite the relatively short time such a star spends with cool supergiant winds, they are the major contributors to mass loss. The value of any bar with width below 0.02~Myr or 0.02~$M_\odot$ is not shown.
}
\label{fig:20_diag}
\end{figure}

Figure~\ref{fig:20_HR} shows the \gls{HR} evolution of each track, while Figure~\ref{fig:20_diag} displays the contribution of each mass loss scheme for a specific model, both in terms of the cumulative time a star spends in each wind phase, and in terms of the cumulative mass that was lost due to specific mass-loss rates. Different mass-loss rates impact differently the core size of the star, leading to different nuclear timescales (Figure~\ref{fig:20_diag}).

\paragraph{\textbf{Optically thin winds}}

These stars spend the vast majority of their \gls{MS} under optically thin winds. In the FESc, MSc and $X_{\rm surf}$ models, the star transitions to a regime with $\log g<3$ (and thus to V01) near \gls{TAMS}. Despite covering the vast majority of the star's lifetime, GM23 accounts for only between 9.2\% and 10.8\% of the total mass lost through winds across all models, with K24 and P25 accounting for even longer evolutionary timescales, but also substantially lower cumulative mass loss. In contrast, while the star remains in the V01 regime for a much shorter time (0.49-0.69~Myr), its mass loss leads to a total mass ejection similar to that occurring during the GM23 phase, but within roughly a 20 times-shorter timescale. Similarly, K24 and P25, being even weaker than GM23, result in less angular momentum ejection and larger cores. This means extended optically thin phases and higher luminosity levels.

\paragraph{\textbf{Cool supergiant winds}}

For these stars, cool supergiant winds represent the dominant contributor to mass loss in isolation. In most models, a 20~$M_\odot$ star typically spends approximately 0.5~Myr in this phase, losing between 4.18 and 7.61~$M_\odot$, which accounts for at least 70\% of its total mass loss. Despite dJ88 being initiated at different effective temperatures ($T_{\rm eff}<$\,10~kK  for the FESc model and $T_{\rm eff}<$\,4~kK  for FESc$_{\rm RSG}$), the star appears to spend a similar duration in the cool supergiant wind phase. This is because the star undergoes a rapid expansion during its \gls{HG} phase between 10~kK and 4~kK (\gls{YSG} regime), consistent with findings by \cite{Zapartas_2025}. However, our models consistently indicate that for these stars strong cool supergiant winds during the \gls{YSG} stage and strong optically thin winds are needed to transition into a \gls{WR} phase. 

\paragraph{\textbf{Optically thick winds}}
Although the FESc, MSc, FESc$_{\rm noHpoor}$, Dutch, Dutch-ish, and $X_{\rm surf}$ models display a final \gls{WR} stage, our models suggest that this phase is relatively short. However, if a 20~$M_\odot$ star does become a \gls{WR} star, the type of transition to thick winds greatly alters what kind of \gls{WR} star it can become (see Figure~\ref{fig:20_HR}), with the $\eta$ transition in the MSc models considerably anticipating the initiation of thick winds past the bistability jump. In the case of the KABS and Warm\_$\Gamma$ schemes, instead, the weak thin winds lead to a slow ejection of angular momentum during \gls{MS}. This leads the star to reach around \gls{TAMS} near-critical rotation velocity on its surface due to efficient angular momentum transport. This additional mechanical mass loss drives the growth of the $\eta$ factor, considerably anticipating the initiation of optically thick winds.

\paragraph{\textbf{Non-WR helium star formation?}} With FESc$_{\rm A24}$, by having V01 in the thin winds phase and dJ88 between 10 and 4~kK, the star loses enough envelope mass to transform into a \gls{WR} star after its \gls{RSG} phase, but only after the star had GM23 and V01 during its late-\gls{CHeB}. At 10.93~Myr ($\sim$99.5\% of its total lifetime), the star reaches $X_{\rm surf}$\,$<$\,0.4, but with still $\Gamma_{\rm e}$\,$<$\,0.5 . This means that the star evolves into a non-\gls{WR} helium star and follows V17. Only at roughly 10.99~Myr ($\sim$99.9\% of its total lifetime) the star meets the $\Gamma_{\rm e}$\,$\geq$\,0.5 transition and becomes a \gls{WR} star following B20. Rather than showing a pathway for non-\gls{WR} helium star formation, these results suggest that the $\Gamma_{\rm e}$ transition, which was originally calibrated in \citet{Bestenlehner_2020} for the formation of WNh stars from \glspl{VMS}, may fail in a narrow parameter space for lower-mass stars.

\subsubsection{The cool supergiant bifurcation: the 40~$M_\odot$ case}
\label{sec:40Msun}

\begin{figure}[!h]
\centering
\includegraphics[width=0.415\textwidth]{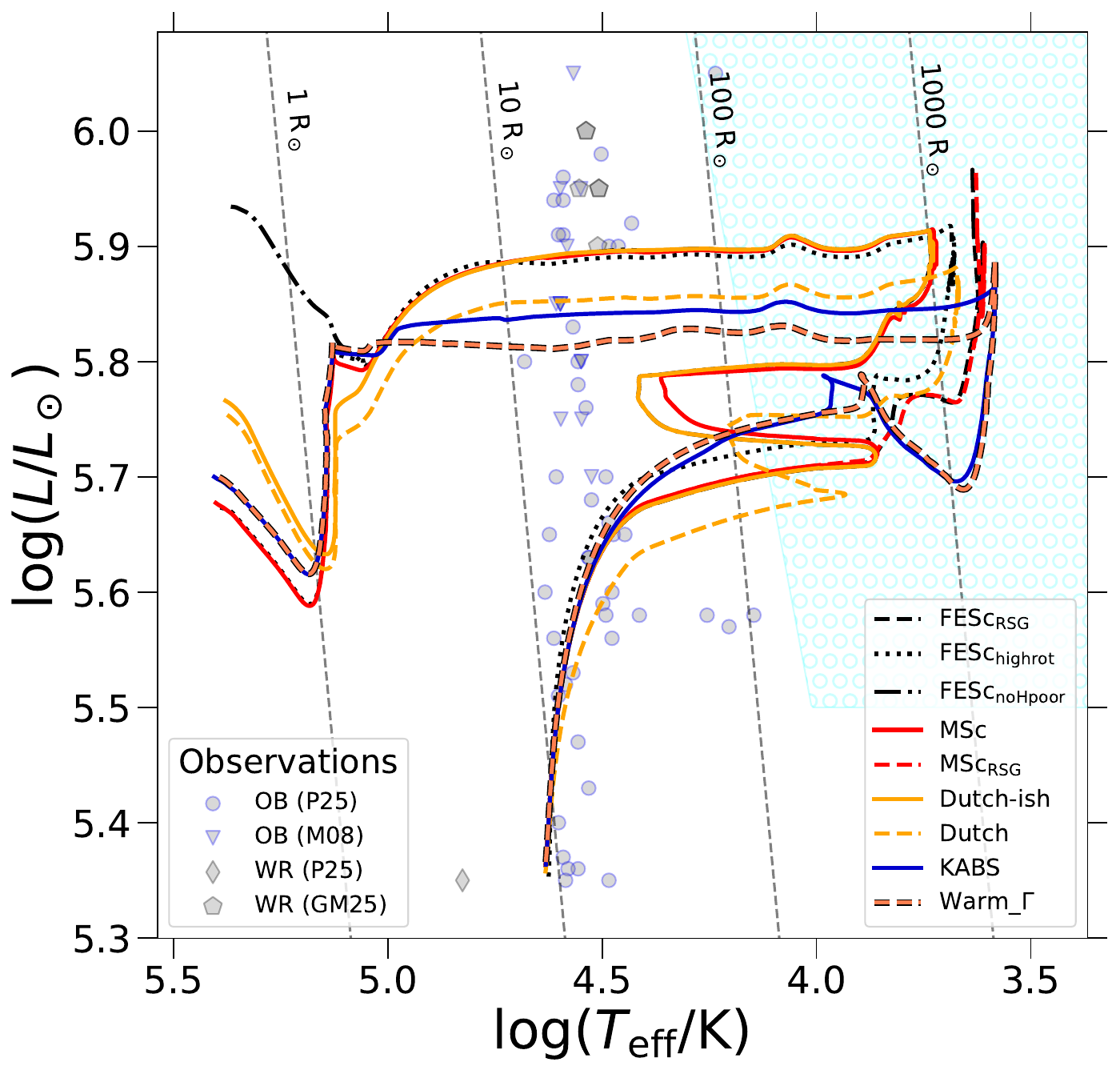}
\caption{HR diagram for $M_{\rm ZAMS}$\,=\,40~$M_\odot$ stars at $Z_\odot$. Submodels that do not show enough variability to be recognizable in the \gls{HR} diagram are not shown. The light blue region represents the \gls{LBV} area past the \gls{HD} limit. The major determinants at this mass for \gls{HR} evolution are the transition condition to cool supergiant winds, and optically thin and thick mass loss.}
\label{fig:40_HR}
\end{figure}

\begin{figure}
\centering
\includegraphics[width=0.455\textwidth]{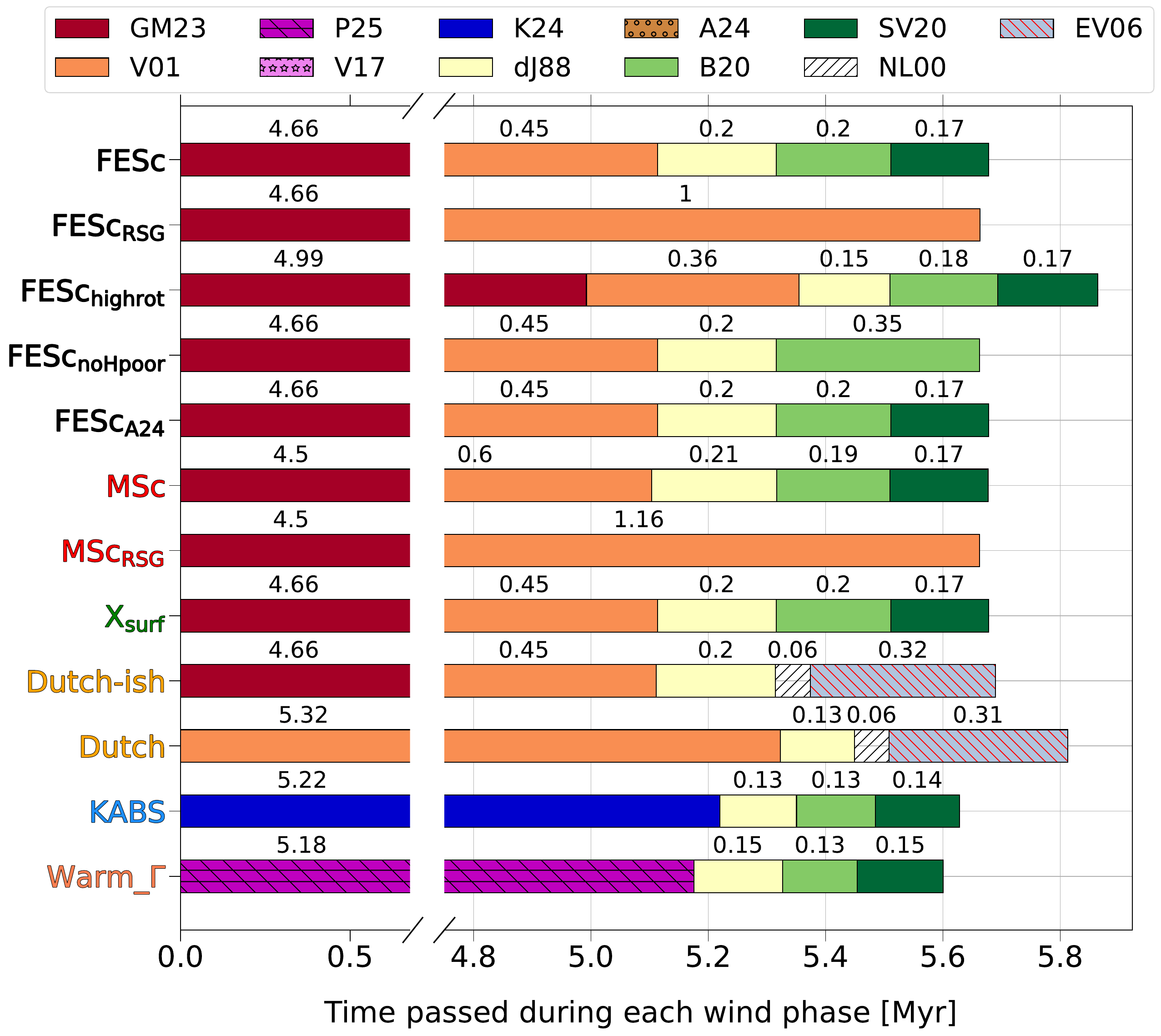}\\
\includegraphics[width=0.44\textwidth]{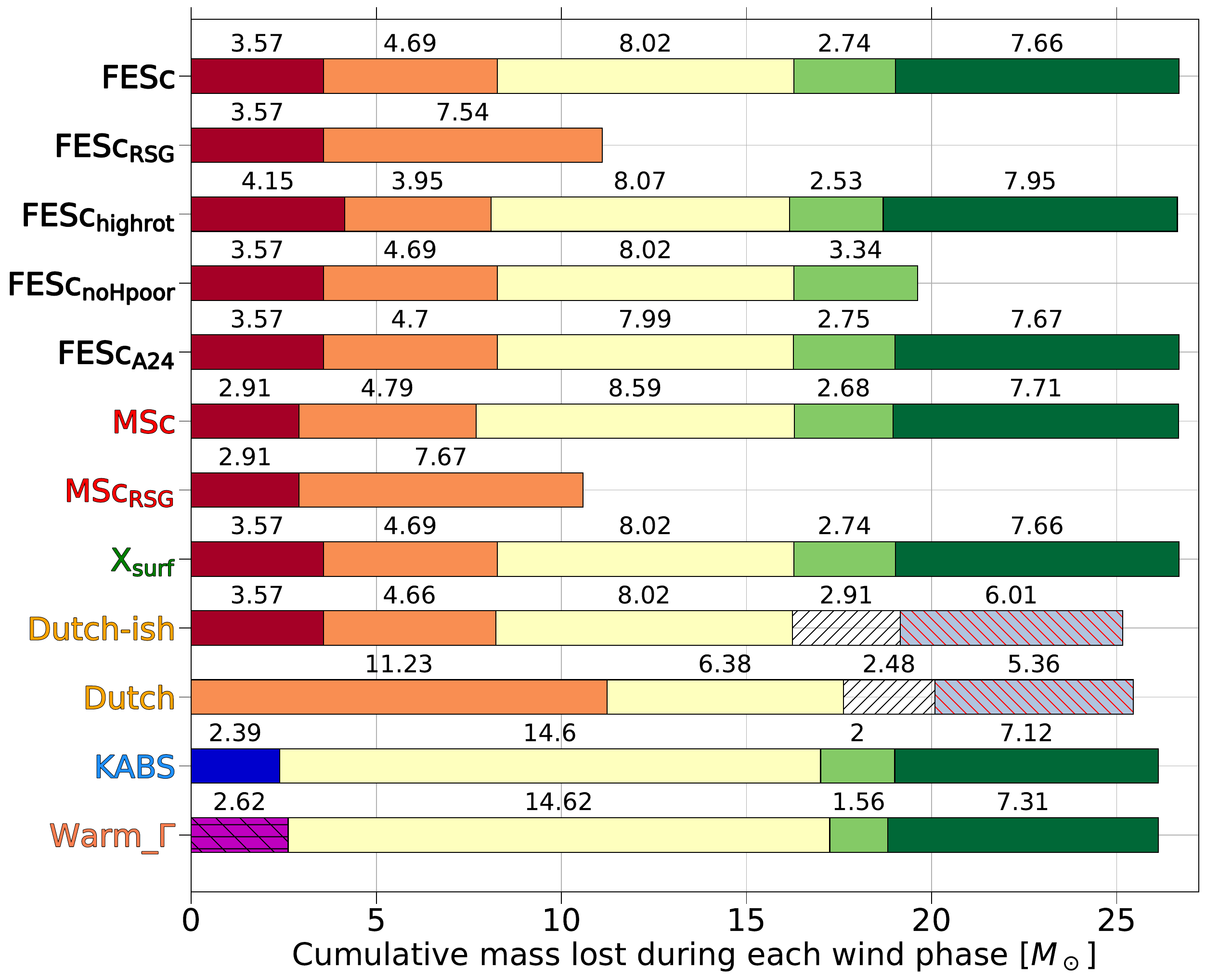}
\caption{Cumulative time and mass loss from each wind scheme for a $M_{\rm ZAMS}$\,=\,40~$M_\odot$ star at $Z_\odot$, according to each proposed evolutionary model.
}
\label{fig:40_diag}
\end{figure}

Figure~\ref{fig:40_HR} shows the \gls{HR} evolution of a 40~$M_\odot$ star following our evolutionary models. Figure~\ref{fig:40_diag}, instead, displays the amount of time and the respective mass that a star loses in each wind phase.

\paragraph{\textbf{Optically thin winds}} Most models show similar evolution for the optically thin wind phase. The reason why FESc and MSc spend different time under GM23 and V01 is due to the fact that MSc transitions earlier to cool \gls{WR} winds. Additionally, increasing the initial rotation (FESc$_{\rm highrot}$) increases the \gls{MS} lifetime and luminosity due to more efficient rotational mixing, and leads to a more compact structure for the star. A similar effect is reached also by weaker optically thin winds such as K24 and P25, which lead to higher \textit{L} during \gls{MS}. 

\paragraph{\textbf{Cool supergiant winds}} These stars do not cool down enough to become \glspl{RSG} due to efficient mixing and ejection of the outermost H-rich layers \citep{Ekstrom_2025}. Only models that allow for dJ88 for \gls{YSG}s experience a significant cool supergiant mass loss (see more in Section~\ref{subsubsec:40_var}). Either a higher initial rotation (FESc$_{\rm highrot}$) or weaker thin winds (KABS, and Warm\_$\Gamma$) lead the star to evolve faster through the cool supergiant phase, but due to higher \textit{L}, the star is losing considerably more mass in proportion. 
On the other hand stronger optically thin winds (Dutch with V01) lead to considerably lower luminosities, which translate into a lower contribution of cool supergiant winds. 

\paragraph{\textbf{Optically thick winds}}
The submodels that only initiate cool supergiant winds for \gls{RSG}s are not ejecting enough mass to deplete the envelope and transition the star to the \gls{WR} phase (see also Section~\ref{subsubsec:40_var}). In spite of different initial conditions, models with the same \gls{WR} winds parameterization evolve to be nearly identical H-free \glspl{WR} and \gls{BH}s.

\subsubsection{Very massive stars entering thick winds during main sequence: the 75~$M_\odot$ case}
\label{sec:75Msun}

\begin{figure}[!h]
\centering
\includegraphics[width=0.445\textwidth]{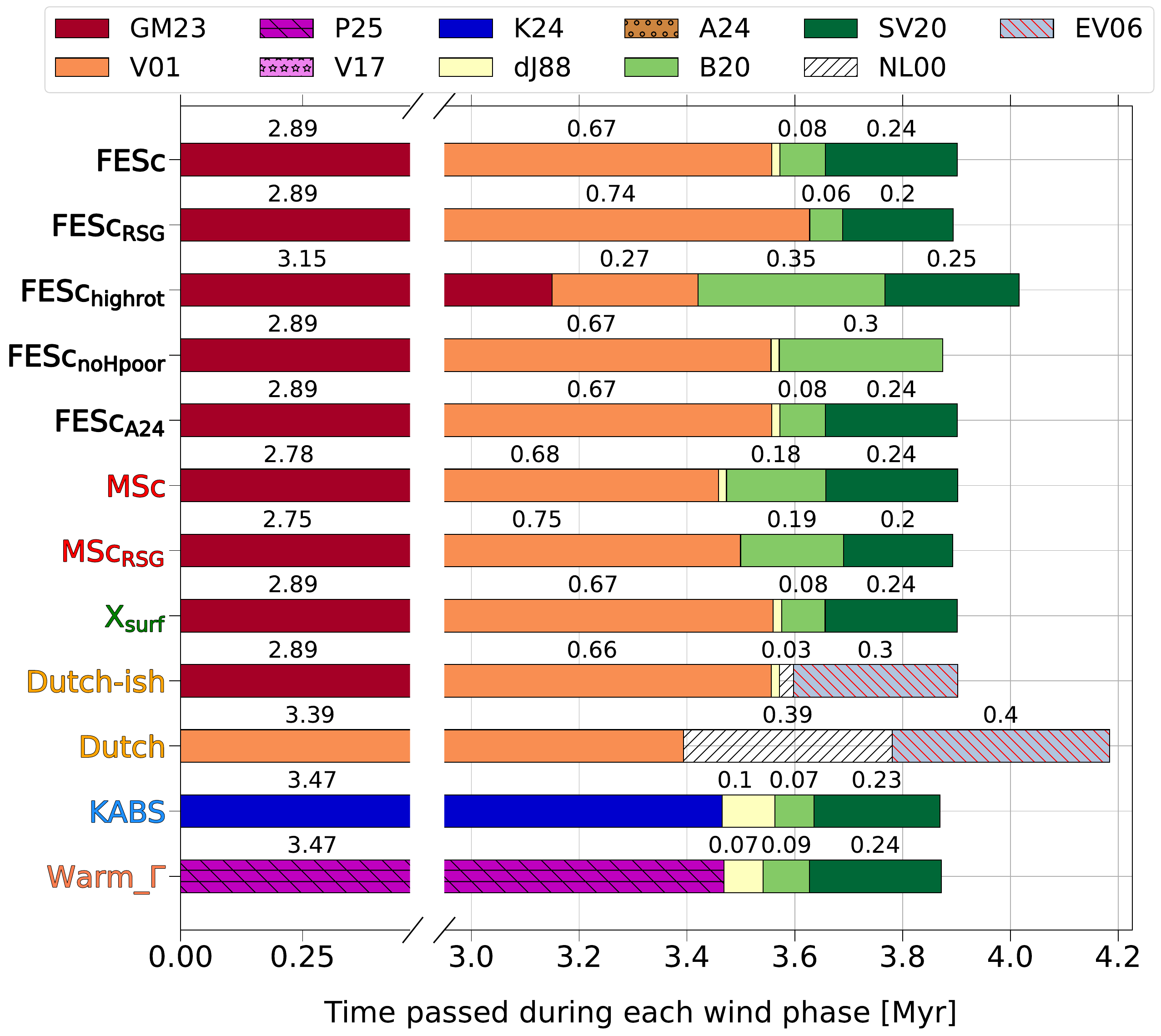}\\
\includegraphics[width=0.44\textwidth]{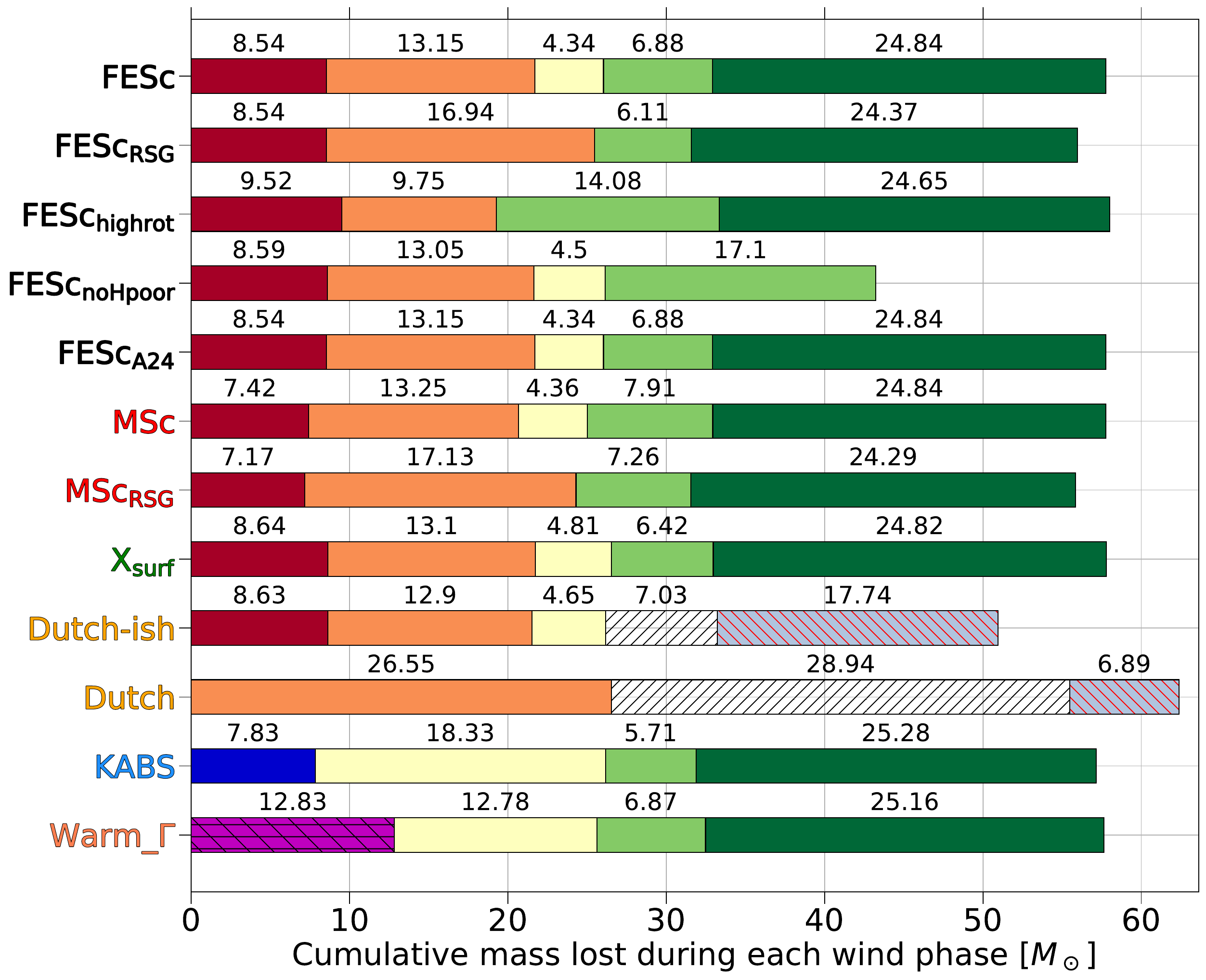}
\caption{Cumulative time and mass loss from each wind scheme for a $M_{\rm ZAMS}$\,=\,75~$M_\odot$ star at $Z_\odot$, according to each proposed evolutionary model. The major contributors are the optically thin and thick winds.
}
\label{fig:75_diag}
\end{figure}

\begin{figure}[!h]
\centering
\includegraphics[width=0.42\textwidth]{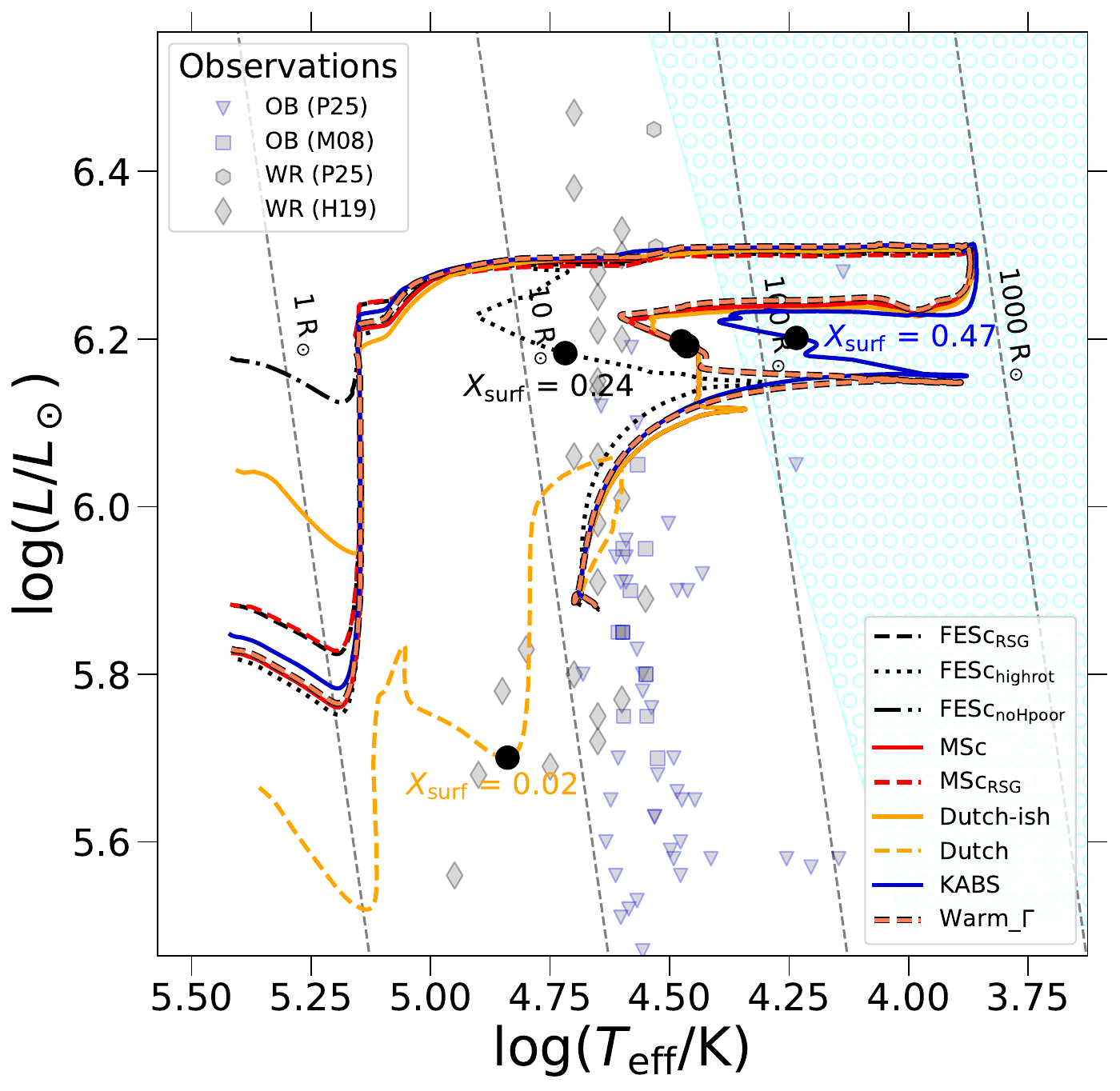}
\caption{HR diagram for $M_{\rm ZAMS}$\,=\,75~$M_\odot$ stars at $Z_\odot$. Black dots represent \gls{TAMS} positions, with the respective surface-H abundances. There is considerable evolutionary variability as a function of different mass-loss rates for optically thin and H-free \gls{WR} winds.}
\label{fig:75_HR}
\end{figure}

Figures~\ref{fig:75_diag} and ~\ref{fig:75_HR} show, according to different models, how much time such a star spends in each mass loss phase, and how much mass it respectively loses. In general, most models lead to quantitatively similar \gls{BH} masses, with the exception of the FESc$_{\rm noHpoor}$, Dutch-ish, and Dutch models. 


\paragraph{\textbf{Optically thin winds}}
For any model incorporating GM23, the star spends roughly 80\% of its lifetime with them. However, GM23 account for a loss of only roughly 7 to 9~$M_\odot$, i.e., between $\sim$\,14\% and 19\% of the total mass lost. In contrast, the star spends roughly a sixth of that time in the V01 phase, but with roughly 1-2.5 times more mass loss. As for the lower-mass cases, the high initial rotation from the FESc$_{\rm highrot}$ model keeps the star compact, with high $\log g$, therefore spending a higher amount of time in its GM23 optically thin phase, but with no significant alteration in terms of total mass loss.

\paragraph{\textbf{Cool supergiant winds}}
The cool supergiant winds phase is only present in models that allow for a transition during the \gls{YSG} phase, and lasts relatively shortly ($\lesssim$\,0.02~Myr), but accounts for a loss of $\sim$\,4 to 6~$M_\odot$. For such stars, cool supergiant winds have a more restricted role if compared to the lower mass case of 20 and 40~$M_\odot$ (up to respectively roughly 8 and 15~$M_\odot$ in mass loss), or no role at all in case of higher rotations (FESc$_{\rm highrot}$). However, the suppression of cool supergiant winds for \gls{YSG}s still leads to the conservation of more envelope mass and a different H-free \gls{WR} evolution (FESc$_{\rm RSG}$ and MSc$_{\rm RSG}$).

\paragraph{\textbf{Optically thick winds}}

Most tracks tend to diverge only near the onset of optically thick winds. For the models implementing GM23 and V01 for thin winds, the $\eta$ transition to optically thick winds is reached 0.2~Myr and 0.8~Myr earlier than respectively the $\Gamma_{\rm e}$ and $X_{\rm surf}$ ones. This difference accounts for at most a 0.3~$M_\odot$ divergence in the final \gls{BH} mass, which we consider negligible. The FESc$_{\rm noHpoor}$ model, which does not have a dedicated model for H-free \gls{WR} winds, produces more massive (14+~$M_\odot$) \gls{BH}s. On the other hand, the Dutch model, incorporating strong V01 and NL00 mass-loss rates, reaches total envelope ejection (i.e. transition to the \gls{WR} phase) already during \gls{MS}, leading to a total mass loss that is significantly higher than other models. Finally, FESc$_{\rm highrot}$ remains more compact than the others due to the retention of more angular momentum. This leads them to enter earlier in the optically thick winds phase, and therefore to have a stronger contribution of \gls{WR} winds to their total evolution. 

\paragraph{\textbf{Cool \gls{WR} winds and the \gls{HD} limit}}

\begin{figure}[!h]
\centering
\includegraphics[width=0.42\textwidth]{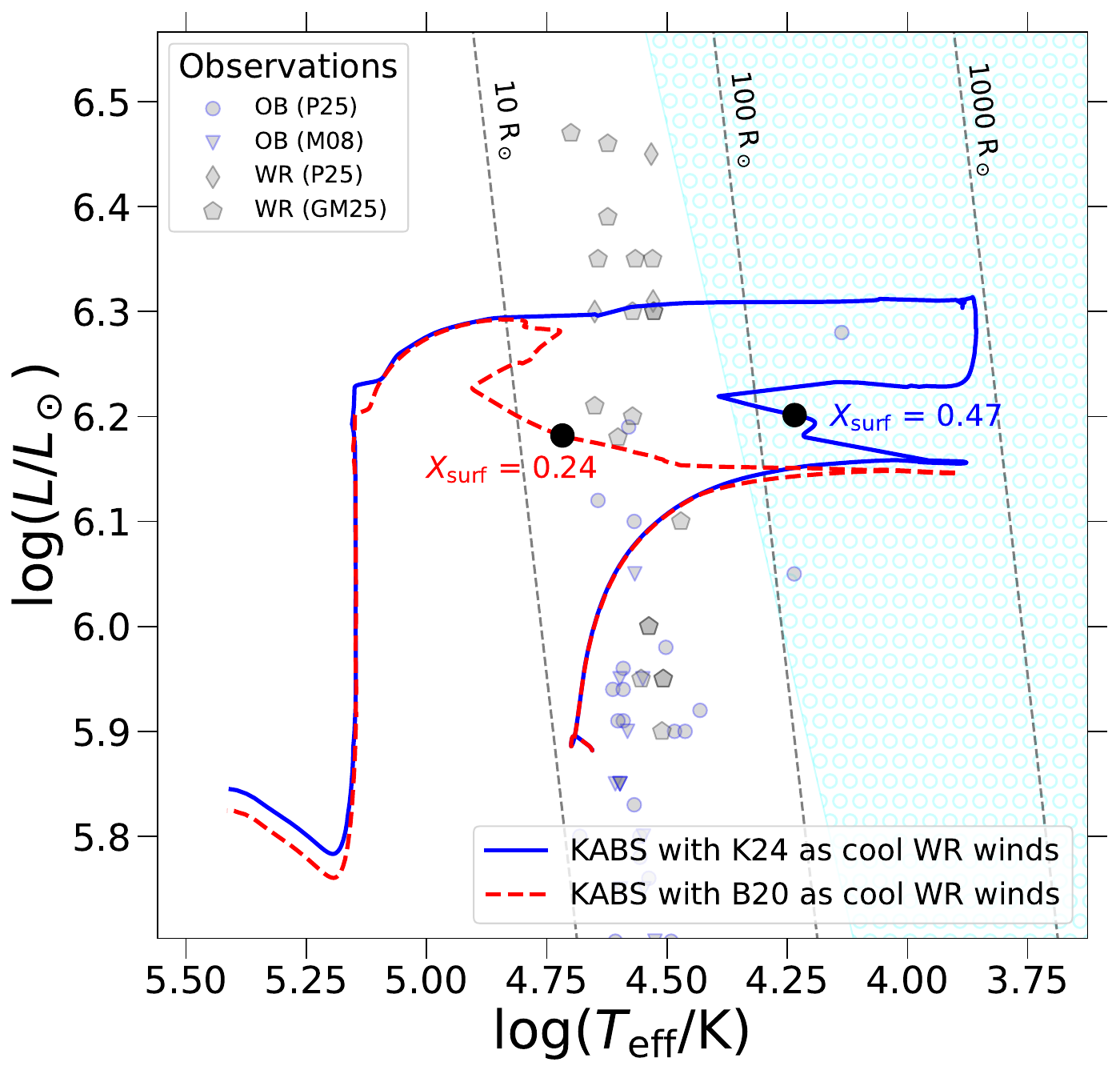}
\caption{HR diagram for $M_{\rm ZAMS}$\,=\,75~$M_\odot$ stars at $Z_\odot$ until the end of core-C burning. KABS model with either K24 (full blue) or B20 (dashed red) for cool \gls{WR} stars ($T_{\rm eff}\leq30$~kK). Black dots represent \gls{TAMS} positions with the respective surface-H abundances, with the light blue representing the \gls{LBV} parameter space beyond the \gls{HD} limit. Using strong B20 winds for the cool \gls{WR} phase evolves the star in near-chemically homogeneous conditions, almost completely respecting the \gls{HD} limit.}
\label{fig:75_coolWR_HR}
\end{figure}

Figure~\ref{fig:75_coolWR_HR} shows the \gls{HR} evolution of a 75~$M_\odot$ star with the KABS model, with either optically thin (K24) or strong optically thick (B20) mass-loss rates for the cool \gls{WR} regime at $T_{\rm eff}\leq30$~kK. Using B20 leads the star to shed its H-rich envelope during \gls{MS} and to reach \gls{TAMS} at $X_{\rm surf}\approx$\,0.24. This makes the star evolve in quasi-chemically homogeneous conditions without becoming a \gls{YSG}, hence respecting the \gls{HD} limit.
Since currently there is no conclusive evidence on which of the two regimes such objects should adhere to \AtlasIp, this represents an important bifurcation point that does more accurately respect the observational lack of luminous cool supergiants without adopting \gls{LBV} models for erupting mass-loss ejections.

\subsubsection{Stars born as Wolf-Rayet objects: 
the 300~$M_\odot$ case}
\label{subsubsec:300Msun}

\begin{figure}[!h]
\centering
\includegraphics[width=0.42\textwidth]{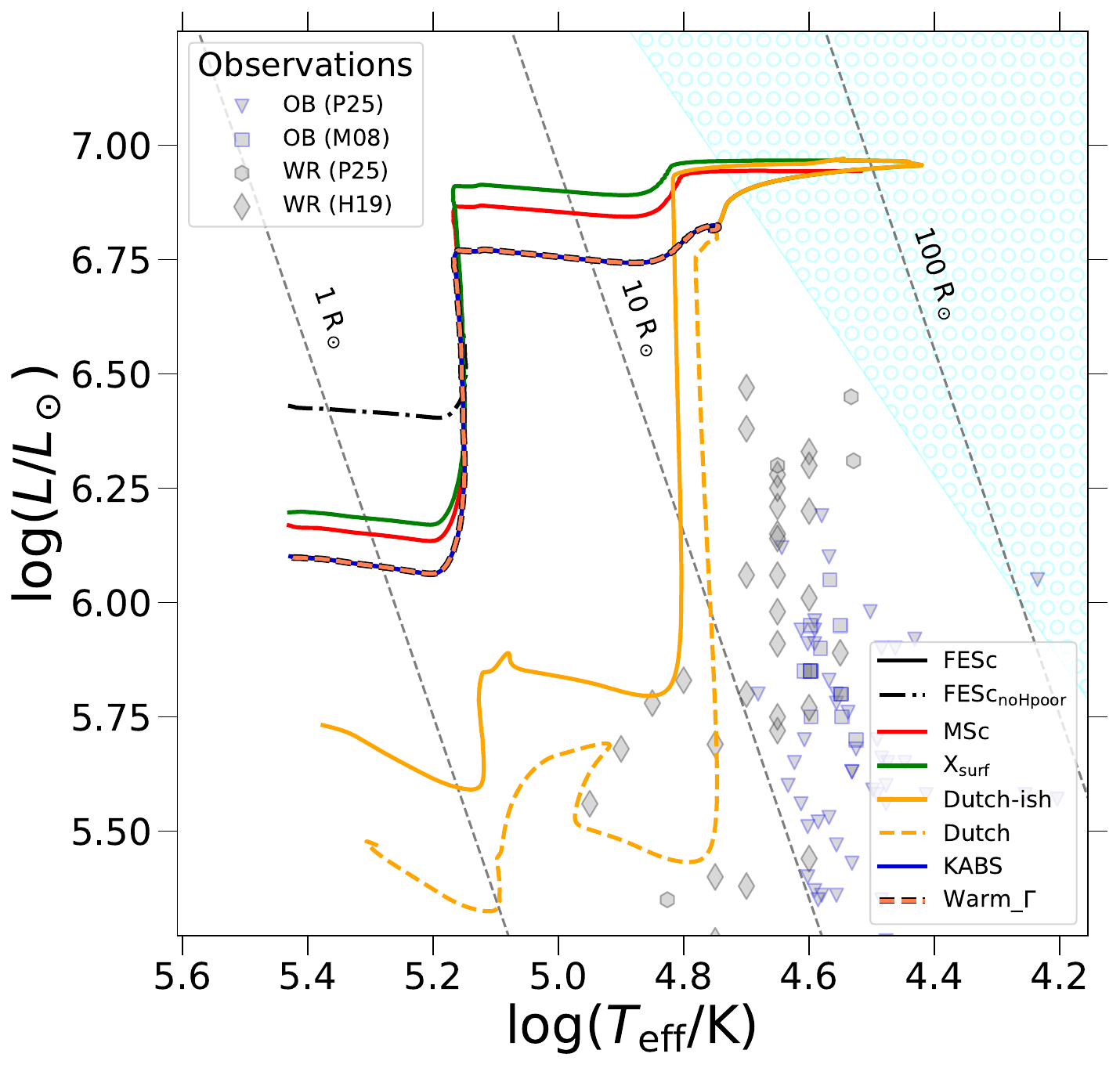}
\caption{HR diagram for $M_{\rm ZAMS}$\,=\,300~$M_\odot$ stars at $Z_\odot$. The colors representing each model remain the same as in Figure
\ref{fig:300_diag}. Models adopting the $\Gamma_{\rm e}$ transition to thick winds shows the most limited increase in radius for such a star. The \gls{LBV} region beyond the \gls{HD} limit is shown in cyan.}
\label{fig:300_HR}
\end{figure}

\begin{figure}[!h]
\centering
\includegraphics[width=0.485\textwidth]{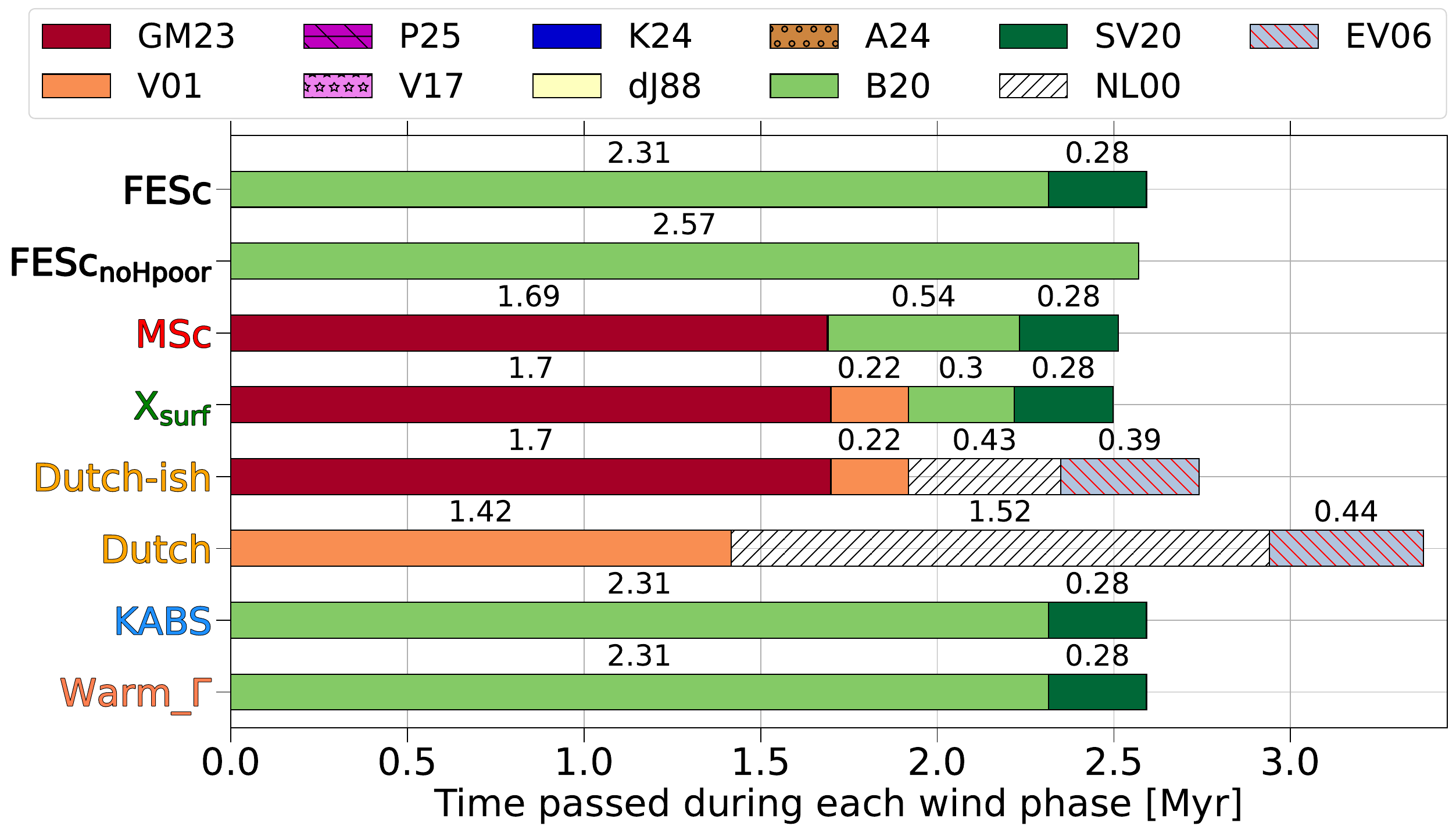}\\
\includegraphics[width=0.485\textwidth]{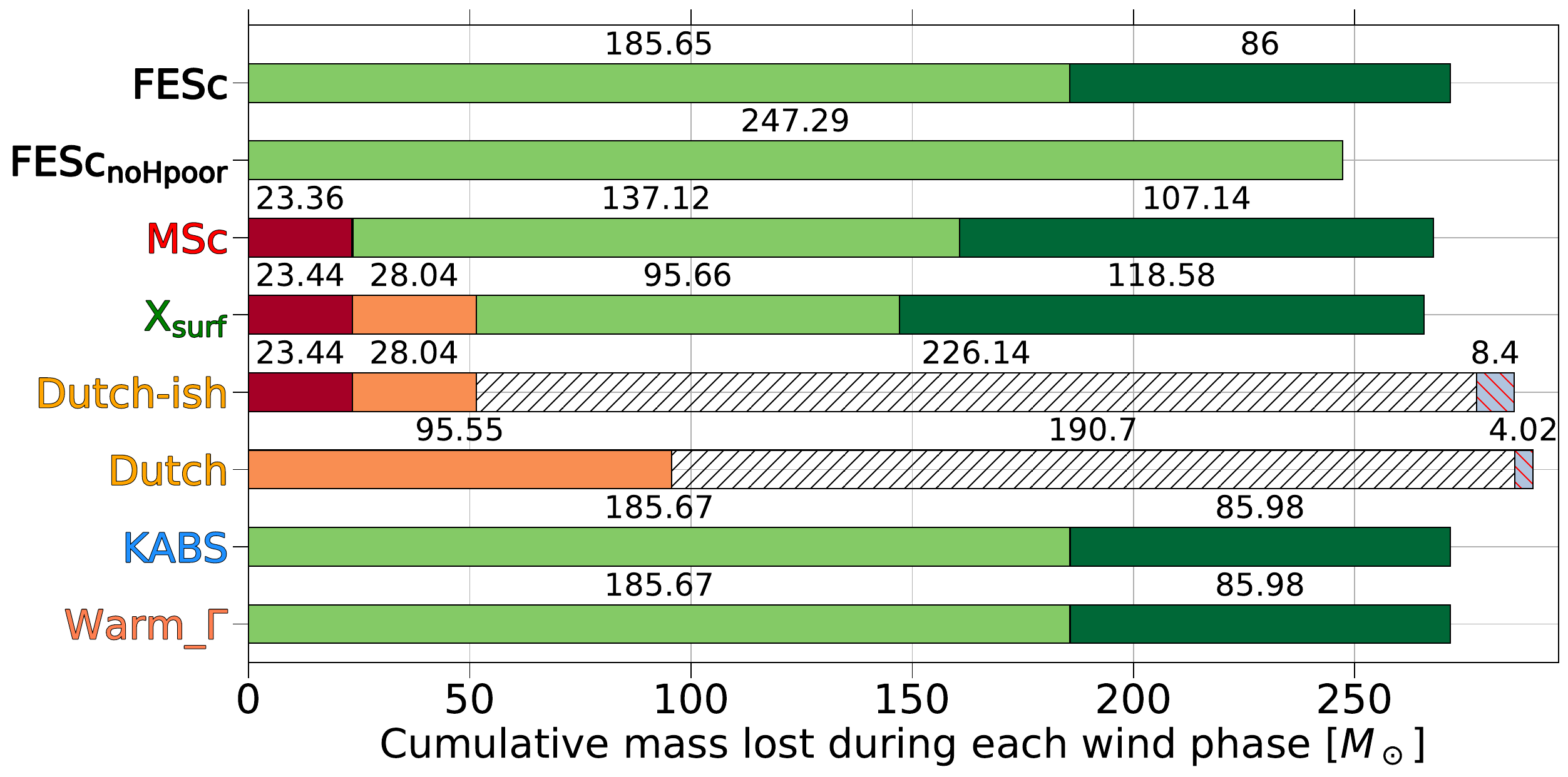}
\caption{Cumulative time and mass loss for a $M_{\rm ZAMS}$\,=\,300~$M_\odot$ star. The majority of mass loss is concentrated to the thick winds phase, with no substantial differences in total mass loss from different transition conditions to thick winds, and no contribution from cool supergiant winds. 
}
\label{fig:300_diag}
\end{figure}

This mass regime exhibits no contribution from cool supergiant winds, as these stars remain relatively compact and evolve directly as \gls{WR}. Figure~\ref{fig:300_HR} shows their \gls{HR} evolution, while Figure~\ref{fig:300_diag} the cumulative time a star spends in each wind phase and the relative mass loss attributed to each scheme.

\paragraph{\textbf{Stars born as WNh}}

In the FESc, KABS and Warm\_$\Gamma$ models, intense luminosity causes stars to be born as WNh objects at \gls{ZAMS}, with their mass loss following exclusively \gls{WR} winds. Also, models that combine strong V01 winds with the $\eta$ transition (see Section~\ref{subsec:thick_trans}), similarly predict the formation of WNh stars near \gls{ZAMS}.

\paragraph{\textbf{Stars born as O-type}}
Models with weak thin winds and either an $\eta$ or $X_{\rm surf}$ transition enter the \gls{WR} phase near \gls{TAMS}, therefore predicting a noticeable contribution of thin winds in the evolution of such a star. For the MSc models, they transition at $\eta\approx0.56$ and $X_{\rm core}\approx$~0.16, while the models with a $X_{\rm surf}$ condition, transition at $X_{\rm core}\approx$~0.10. In particular, models with the condition $X_{\rm surf}$ display an expansion beyond 100~$R_\odot$. This comes with the exception of the Dutch model, which transitions to thick winds halfway in its \gls{MS} due to strong V01 winds, without any noticeable expansion.




\section{Discussion}
\label{sec:discussion_uncertainties}

Our results are subject to several key physical uncertainties inherent in 1D stellar evolution on mass-loss rates and transition conditions \AtlasIp.

Forthcoming data from \gls{LVK} O4, Gaia DR4 \citep{GaiaBH3}, and LSST \citep{Ivezic_2019} demand a deeper understanding of stellar evolution models. We must not only constrain uncertainties in massive star evolution but also systematically map how models respond to chosen initial conditions and physical assumptions. Success in this area is crucial for identifying common patterns across theoretical frameworks, improving model control, and ultimately making robust connections between a \gls{BH} and its stellar progenitor.

\subsection{Fundamental drivers in BH mass predictions}

\begin{figure*}[!ht]
\centering
\includegraphics[width=1\textwidth]{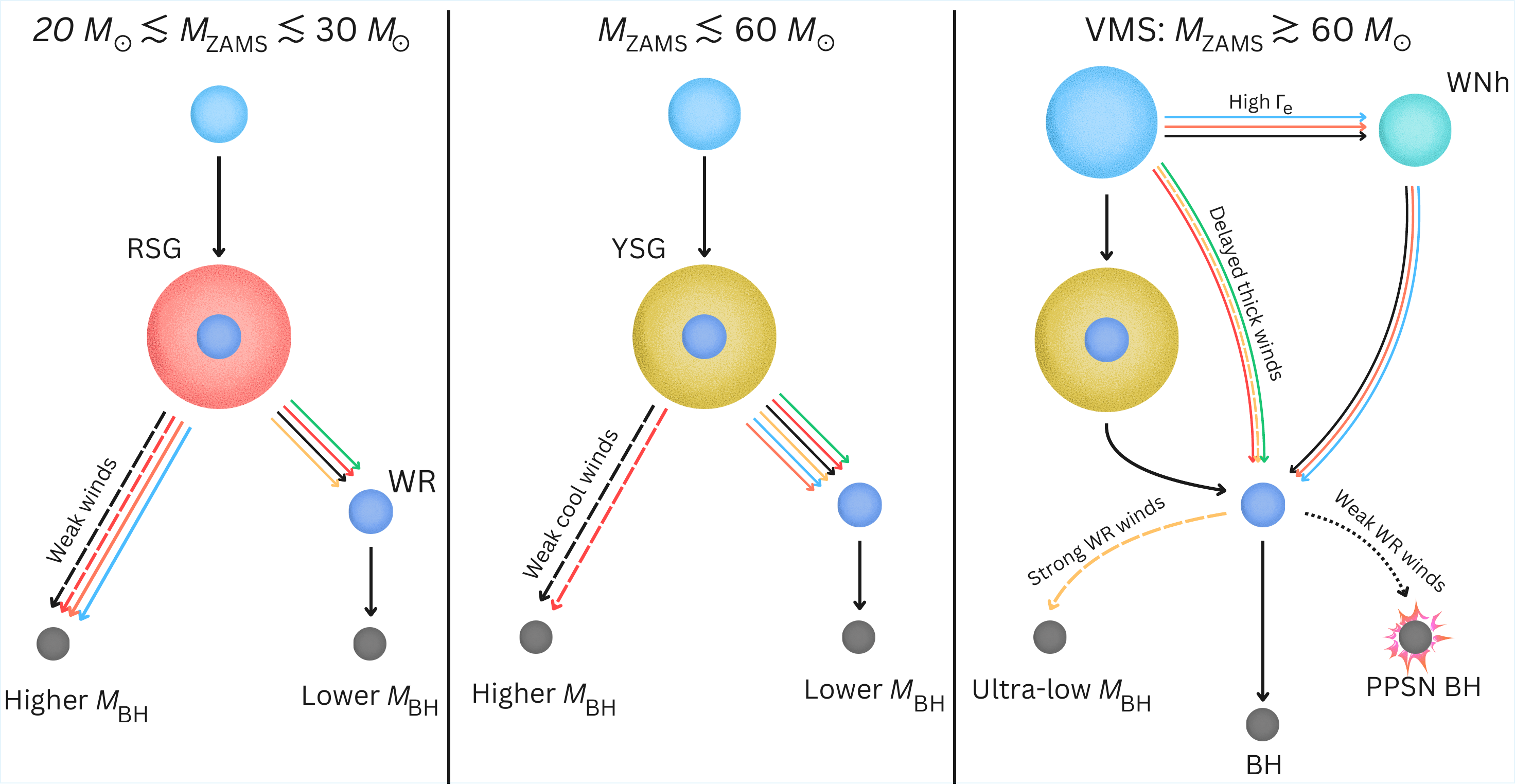}
\caption{The main evolutionary channels for \gls{BH} progenitors. $M_{\rm BH}$ is determined by a key bifurcation: stars that collapse as cool supergiants form higher-mass \gls{BH}s, while those that evolve as \gls{WR} stars lose more mass and form lower-mass remnants. The full single black line represents a generic behavior, while all the other lines represent the respective models in Figure~\ref{fig:BHmass}.
}
\label{fig:st_to_BH_evo}
\end{figure*}

Our analysis reveals that the \gls{BH} mass at $Z_\odot$ is primarily controlled by a fundamental bifurcation in its progenitor's evolution, as shown in Figure~\ref{fig:st_to_BH_evo}. The key determinant is whether the star successfully becomes a \gls{WR} star before core-collapse. This single outcome, which recent literature suggest to be dominated by winds rather than mass transfer events in binaries \citep{Shenar_2020a}, dictates which physical processes will dominate the final stages of mass loss and the \gls{BH} mass.

If a star enters the \gls{WR} phase, its final mass is overwhelmingly set by the strength of its optically thick winds. These winds are so powerful that they largely erase the signature of the preceding mass-loss history from the optically thin and cool supergiant phases. Across our models, for a given thick wind prescription, the final \gls{BH} masses nearly converge. Despite the optically thin mass loss during \gls{MS} can affect the length of the \gls{WR} phase \citep{Josiek_2024}, we find that this variability, nor the one arising from different thick winds transition conditions, does not heavily influence \gls{BH} masses. 

On the other hand, if a star fails to shed its envelope, it collapses as a cool supergiant. The final \gls{BH} is then formed from the whole stellar core mass plus at least a fraction of the retained envelope mass, leading to a substantially more massive remnant. This mechanism is the primary driver for the prominent \gls{BH} mass peaks that appear around $M_{\rm ZAMS}$\,$\sim$\,40~$M_\odot$ in models that limit strong cool supergiant winds (Figure~\ref{fig:BHmass}).

This bifurcation simplifies the challenge of predicting the \gls{BH} mass distribution at $Z_\odot$, since the parameter space of isolated stellar evolution can be reduced to two fundamental bottlenecks in the \gls{BH} population: whether mixing, cool supergiant, and thin winds are strong enough for forming a \gls{WR} star \footnote{We do highlight that, despite our analysis does not focus on eruptive \gls{LBV} mass loss, also this factor could deeply influence the formation of \gls{WR} stars, effectively further increasing the chances of ejecting any H-rich layer.}, and the mass-loss rates of optically thick winds.

\subsection{Observational tests and implications}

The correct identification of \gls{BH}s that come from the collapse of $Z_\odot$ stars can only happen within binary or multiple systems with an active stellar companion. The observation of X-ray binaries is a powerful tool to this goal, but discriminating how much of the observed \gls{BH} mass comes from the collapse of the progenitor star and how much from binary interactions may be challenging (see Section~\ref{subsec:cygX1}). However, astrometric observations of \gls{BH} binaries at wide orbital distances, such as those from Gaia BH1 and BH2 \citep{ElBadry_2023,ElBadry_2023b},may be used to better understand the evolution of massive binaries, constrain highly uncertain binary physics such as natal kicks and mass transfer efficiency, and calibrate evolutionary models \citep{Kruckow_2024, Mapelli_2026}. 


\paragraph{\textbf{Observed $M_{\rm BH}$ distribution}} An abundant population of observed \glspl{BH} between roughly 20 and 30~$M_\odot$, reflecting the more frequent formation of stars at $M_{\rm ZAMS}$\,$<$\,60~$M_\odot$ rather than \glspl{VMS}, may prove the correctness of the MSc$_{\rm RSG}$ and FESc$_{\rm RSG}$ models, and a negligible contribution of both \gls{LBV} and cool supergiant winds in the evolution of \gls{BH} progenitors. On the other hand, a small or absent population of \gls{BH}s within that mass regime may imply that \glspl{VMS} are the only stars that evolve into such massive objects. This would show that non-\glspl{VMS} face intense phases of mass loss at low-$T_{\rm eff}$ levels, be it via cool supergiant winds or \gls{LBV} eruptions.

\paragraph{\textbf{WR vs. OB population}} At a given $M_{\rm ZAMS}$ different models for \gls{VMS}s predict the transition to the \gls{WR} stage at different ages. This means that models predict different likelihoods for whether a star can be observed as an OB or \gls{WR} star. With enough OB and \gls{WR} observations, \gls{VMS} winds can be tested, as we already did in Sections~\ref{subsec:thick_trans} and \ref{subsubsec:300Msun}. A population of OB objects at a given luminosity range would disprove models that lead stars to enter the \gls{WR} phase at \gls{ZAMS}, while a population of mainly \gls{WR} stars would disprove models that predict a long-lasting OB phase within that regime.

\paragraph{\textbf{Disappearing stars}} Beyond the mass of the resulting \gls{BH}, the evolutionary bifurcation between \gls{RSG} and \gls{WR} pathways directly dictates the expected progenitor type. This has significant implications for searches targeting disappearing stars as direct evidence of \gls{BH} formation \citep[e.g.][]{Gilkis_2026}, as the detectability of such an event depends heavily on the H-rich envelope retention.

\paragraph{\textbf{Binaries and dense stellar environments}} FESc$_{\rm noHpoor}$ is the only model in our sample that produces $M_{\rm BH}$\,$\gtrsim$\,30~$M_\odot$ from artificially underestimating mass loss during the H-free \gls{WR} phase. This means that conservatively at $M_{\rm BH}$\,$\gtrsim$\,30~$M_\odot$, we do not expect to find \gls{BH}s that formed from the isolated evolution of their stellar progenitor. Such \gls{BH}s could come from \gls{RLOF} events between a stellar companion and the \gls{BH} itself, a stellar merger that could increase the envelope mass and bypass the \gls{PPSN} mechanism, the dynamical encounters of \gls{BH}s in globular clusters, or the merger of two \gls{BH}s. The detection of a \gls{BH} in $Z_\odot$ environments of such a mass would be a clear indication of non-isolated evolution of the progenitor star.

\section{Conclusions}
\label{conclusions}

The evolution of massive stars and the masses of the black holes they produce are central to modern astrophysics, yet theoretical predictions vary significantly across different studies. This divergence often stems from differing assumptions about stellar winds, with most models developed without systematic, cross-calibrated comparisons. The primary motivation for this study was to address this ambiguity by constructing a ``Wind Atlas'': a comprehensive investigation of a wide range of wind prescriptions within a common evolutionary framework.

Our analysis reveals that the final mass of a black hole at solar metallicity is primarily controlled by a fundamental bifurcation (Figure~\ref{fig:st_to_BH_evo}) in its progenitor's evolution. The key ingredient is whether the star successfully becomes a Wolf-Rayet star before core-collapse.

\begin{itemize}
    \item The Wolf-Rayet Pathway: If a star enters the Wolf-Rayet phase, its final mass is overwhelmingly set by the strength of its optically thick winds. These winds are so powerful that they largely erase the signature of the preceding mass-loss history. Consequently, stars that follow different evolutionary tracks but adopt the same thick wind prescription produce black holes of very similar masses.
    \item The Supergiant Pathway: If a star fails to shed its envelope, it collapses as a cool supergiant. In this scenario, it retains a significant portion of its mass, forming a substantially more massive remnant, despite the loosely-bound H-rich envelope may still be ejected during core-collapse. This mechanism is the primary driver for the prominent black hole mass peaks that appear around an initial mass within 50~$M_\odot$ in some of our models and in the literature.
\end{itemize}

The timing of the transition to strong, optically thick winds also correlates with a key observational benchmark: the Humphrey-Davidson limit. We find an inverse relationship where models that trigger an early transition to strong winds produce compact stars that tend to respect more the Humphrey-Davidson limit. In contrast, models with a delayed onset of strong mass loss allow stars to expand significantly, causing them to evolve beyond this empirical boundary.

These theoretical divergences offer clear, testable predictions. The timing of the transition to the Wolf-Rayet phase, which dictates both the final black hole mass and the stellar \gls{HR} evolution, can be constrained by observing stellar populations in the Milky Way. The predicted Wolf-Rayet to OB star number ratio in a given luminosity range is a direct model-independent probe of this timing. Such observations can therefore be used to challenge models that predict either a very late or a very early transition for massive stars.

\section*{Code availability}
A detailed description of the population synthesis code {\tt StarEstate} \citep{Romagnolo_2025_StarEstate} can be found at \url{https://github.com/AmedeoRom/StarEstate}. Our {\tt MESA} setup can be found at \url{github.com/AmedeoRom/Stellar_Winds_Atlas}.

\section*{Software acknowledgements}
{\small
This work used the following software packages: \texttt{matplotlib} \citep{Hunter:2007}, \texttt{numpy} \citep{numpy}, \texttt{pandas} \citep{mckinney-proc-scipy-2010, pandas_13819579}, \texttt{pasta-marker} \citep{Pasta_2024,Pasta_2025}, and \texttt{python} \citep{python}.
Information about software citation aggregated using \texttt{\href{https://www.tomwagg.com/software-citation-station/}{The Software Citation Station}} \citep{software-citation-station-paper, software-citation-station-zenodo}.}

\vspace*{-0.2cm} 

\section*{Acknowledgements}
Computations for this article have been performed using the computer cluster at CAMK PAN. AR acknowledges the support from the Polish National Science Center (NCN) grant Maestro (2018/30/A/ST9/00050). AR and LMS acknowledge financial support from the European Research Council for the ERC Consolidator grant DEMOBLACK, under contract no. 770017 and from the German Excellence Strategy via the Heidelberg Cluster of Excellence (EXC 2181 - 390900948) STRUCTURES. LMS acknowledges support from the Alexander von Humboldt Foundation. LB acknowledges support by the Deutsche Forschungsgemeinschaft (DFG, German Research Foundation) in the form of a Walter Benjamin position -- Projektnummer 555003977. ACGM thanks the support from project 10108195 MERIT (MSCA-COFUND Horizon Europe). DP acknowledges financial support from the FWO in the form of a junior postdoctoral fellowship No. 1256225N. AG acknowledges support from the Isaac Newton Trust (University of Cambridge). The authors acknowledge the helpful interactions with J. Klencki, A. Sander, M. Renzo, T. Wagg, J. Vink, G. Sabhahit, T. Janka, and T. Shenar .

\bibliography{ms}
\bibliographystyle{aasjournal}


\begin{appendix}

\section{Limitations of using $\Gamma_{\rm e}$ as a high-mass loss proxy}

\begin{figure}[!h]
\centering
\includegraphics[width=0.485\textwidth]{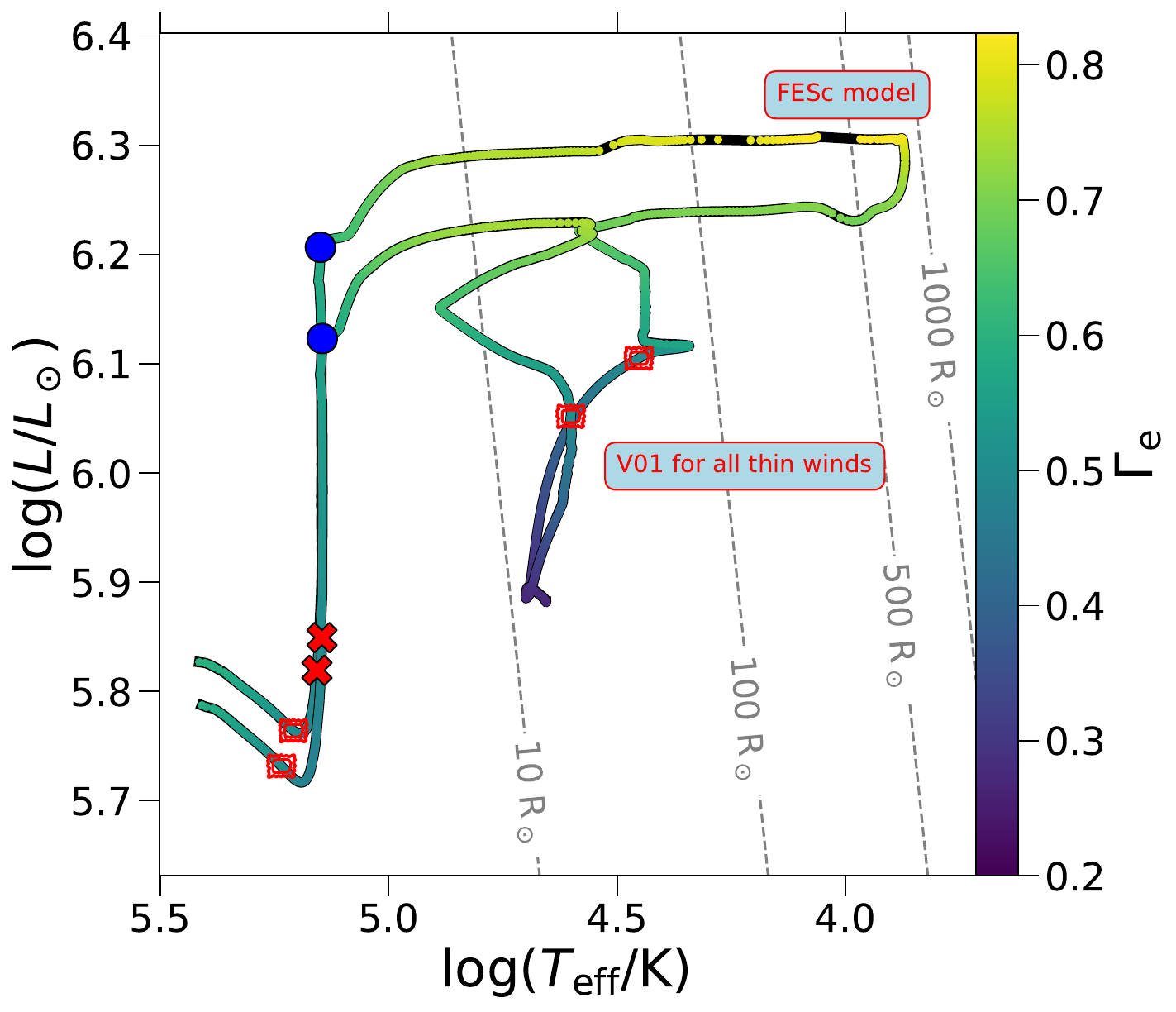}
\caption{HR evolution of a 75~$M_\odot$ star at $Z_\odot$ following the FESc model and the FESc$_{\rm V01}$. The colorbar represents the evolution of $\Gamma_{\rm e}$. Black lines are plotted under the stellar tracks' scatter points to increase contrast. The red ravioli represent where $\Gamma_{\rm e}$ increases past the 0.5 threshold, the red crosses where $\Gamma_{\rm e}$ decreases under 0.5, and the blue dots when H-free winds are initiated. The $\Gamma_{\rm e}$ value goes briefly below the 0.5 threshold during the H-free \gls{WR} phase.}
\label{fig:75_lessGamma}
\end{figure}

Figure~\ref{fig:75_lessGamma} shows the \gls{HR} and $\Gamma_{\rm e}$ evolution of a 75~$M_\odot$ star at $Z_\odot$ following the FESc and FESc$_{\rm V01}$ models (for more information on the evolutionary diagnostics of a 75~$M_\odot$ star, we refer to Section~\ref{sec:75Msun}).

\paragraph{\textbf{FESc models}}
While the $\Gamma_{\rm e}$\,$>$\,0.5 condition is a robust, model-independent indicator for the transition of OB stars into WN stars, it has a key limitation during the subsequent evolution into the H-free phase. Our models show that $\Gamma_{\rm e}$ can drop back below the 0.5 threshold for H-free \gls{WR} stars, even as the star physically maintains its optically thick winds. This dip is caused solely by a sharp drop in luminosity that occurs when the star becomes H-free (indicated by the blue dots in Figure~\ref{fig:75_lessGamma}) and readjusts to maintain thermal balance due its low opacity. This low-$\Gamma_{\rm e}$ period is brief, lasting $\sim$0.08~Myr before $\Gamma_{\rm e}$ goes back above the 0.5 threshold. A strict application of the $\Gamma_{\rm e}$ criterion would therefore incorrectly re-classify the object as no longer having optically thick winds. To address this, our models treat the $\Gamma_{\rm e}$\,$>$\,0.5 condition as a one-way switch for initiating the \gls{WR} phase. Once a star meets this criterion, it continues to be treated as a \gls{WR} star for the remainder of its evolution, ensuring the appropriate mass loss scheme is used regardless of the subsequent value of $\Gamma_{\rm e}$. Ultimately, this temporary deviation occurs in a very narrow parameter space, highlighting that the $\Gamma_{\rm e}$ criterion may still be used to model the \gls{WR} phase.

\paragraph{\textbf{MSc and $\eta$-based models}}

In our MSc models we do not change the H-rich optically thick winds of FESc, i.e. V01 for cool \gls{WR}s, and B20 for warmer winds. Despite in the original \cite{Sabhahit_2022} models stars were not shown to have a considerably decreasing trend for $\Gamma_{\rm e}$ that would switch from stronger (V11) to weaker (V01) mass-loss rates, we show that this might not be an universally valid behavior. Depending on initial conditions such as the adopted stellar winds, mass loss may oscillate in intensity depending on the $\Gamma_{\rm e}$ and mass-loss histories.

\section{Mass-loss history}

\begin{figure*}[!ht]
\centering
\includegraphics[width=0.30\textwidth]{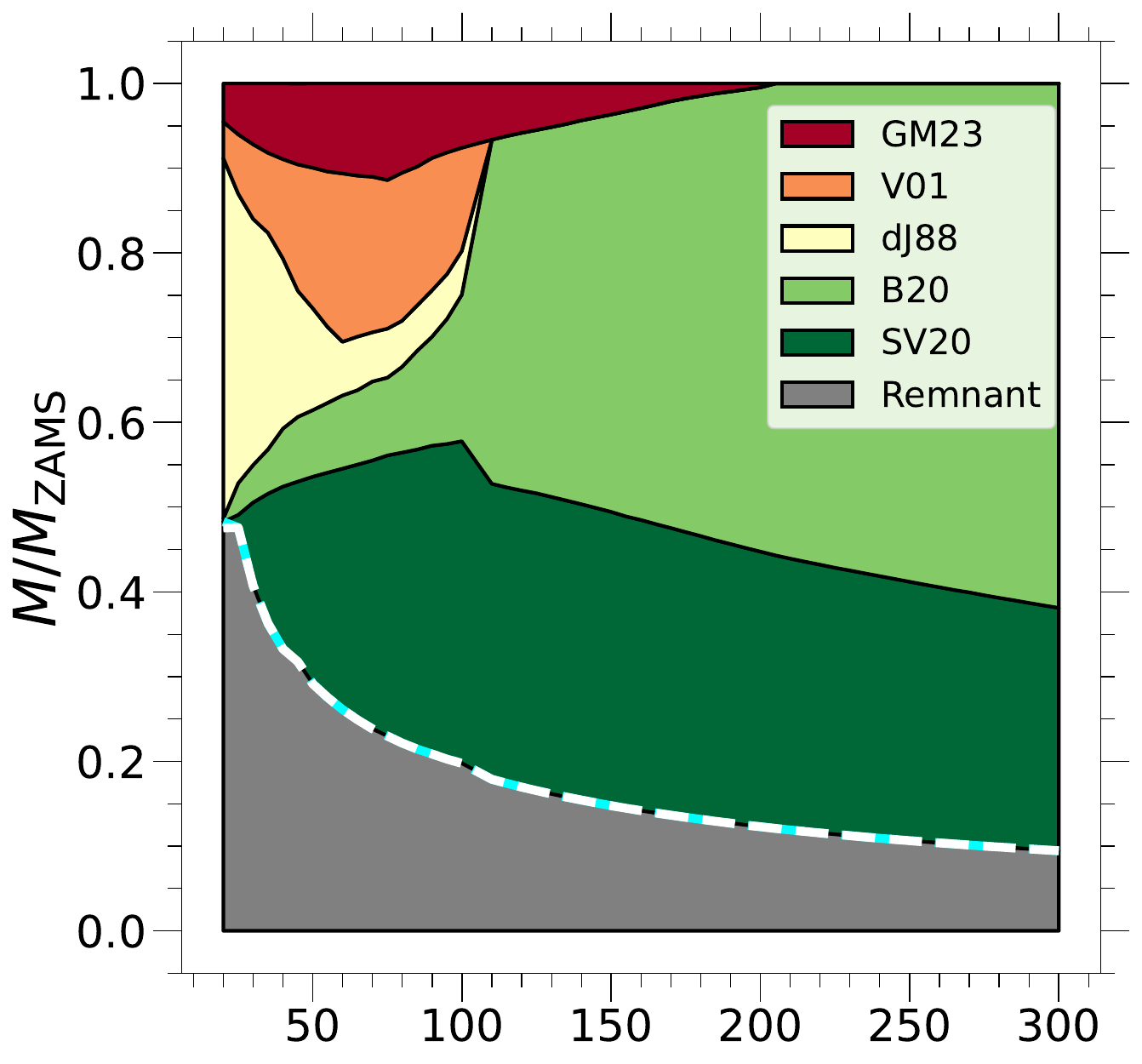}
\includegraphics[width=0.28\textwidth]{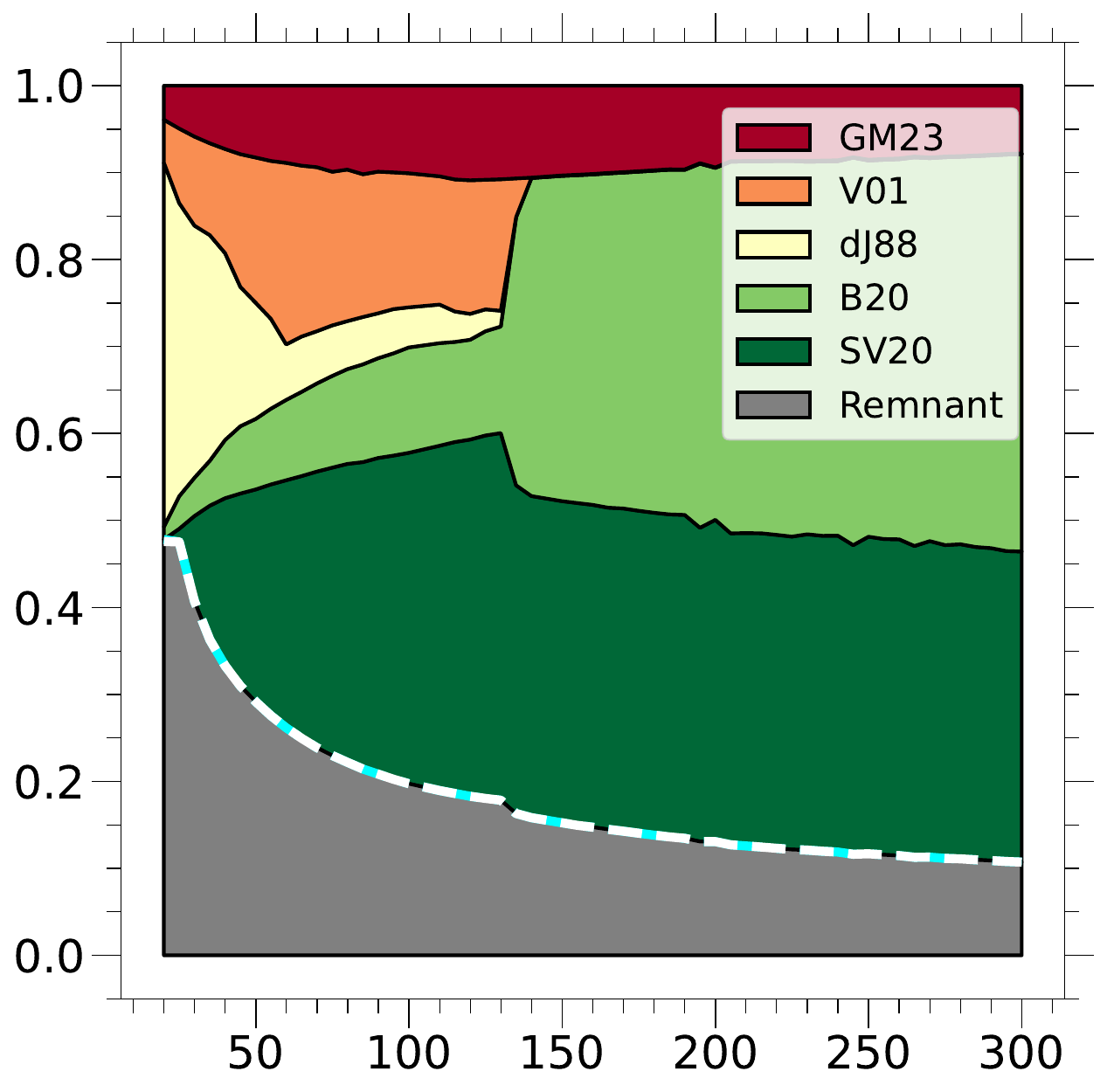}
\includegraphics[width=0.28\textwidth]{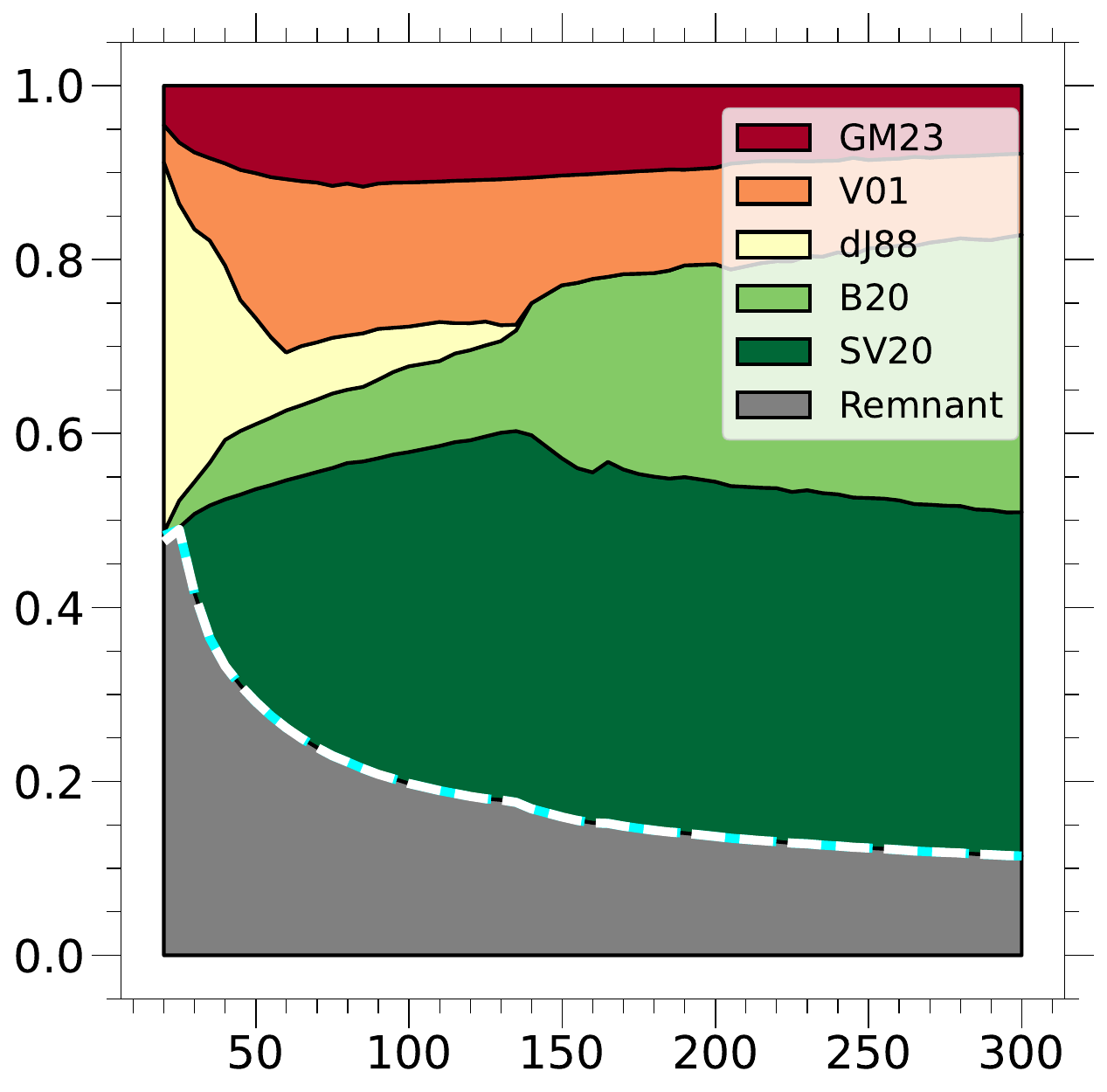}\\
\includegraphics[width=0.30\textwidth]{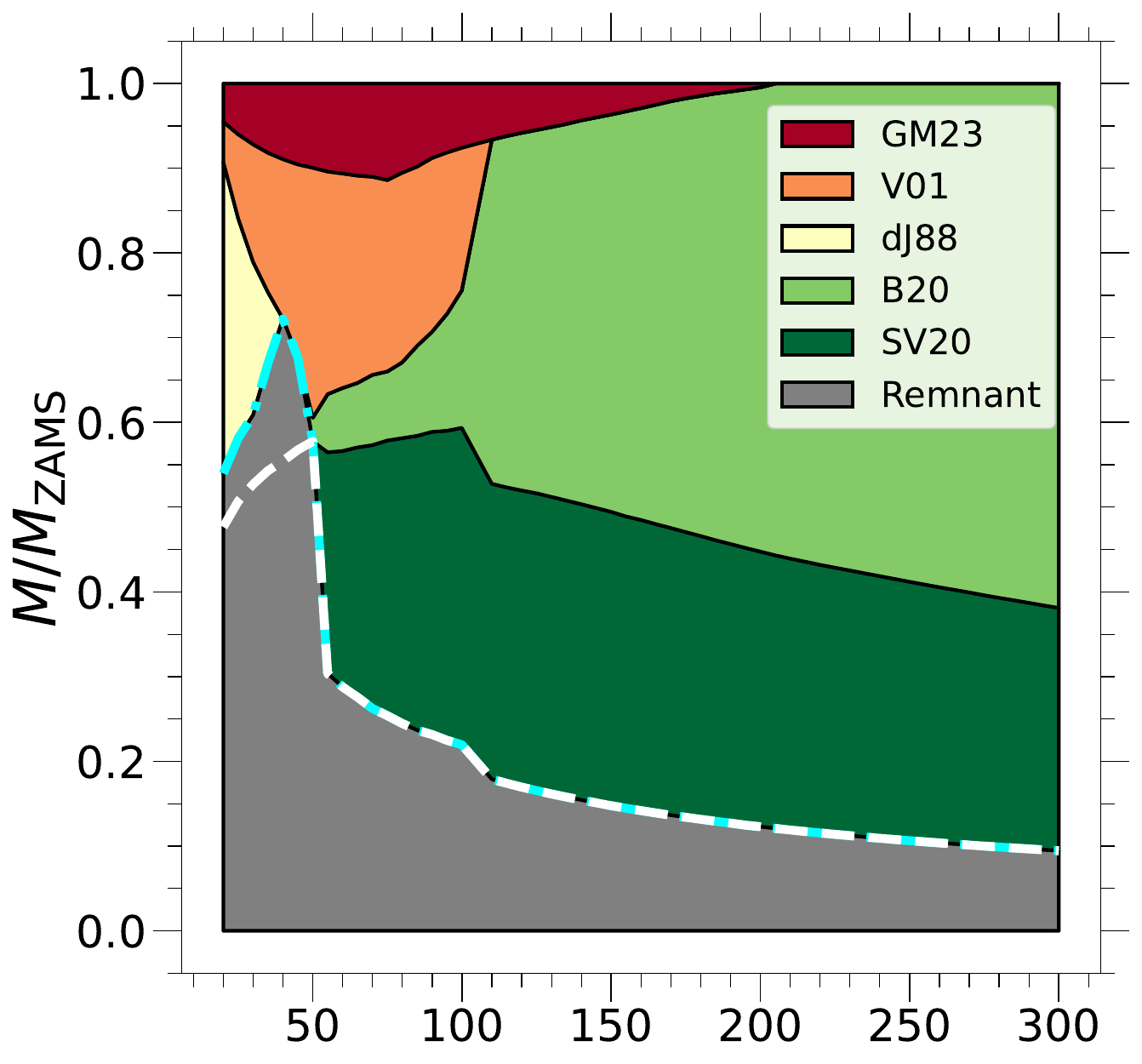}
\includegraphics[width=0.28\textwidth]{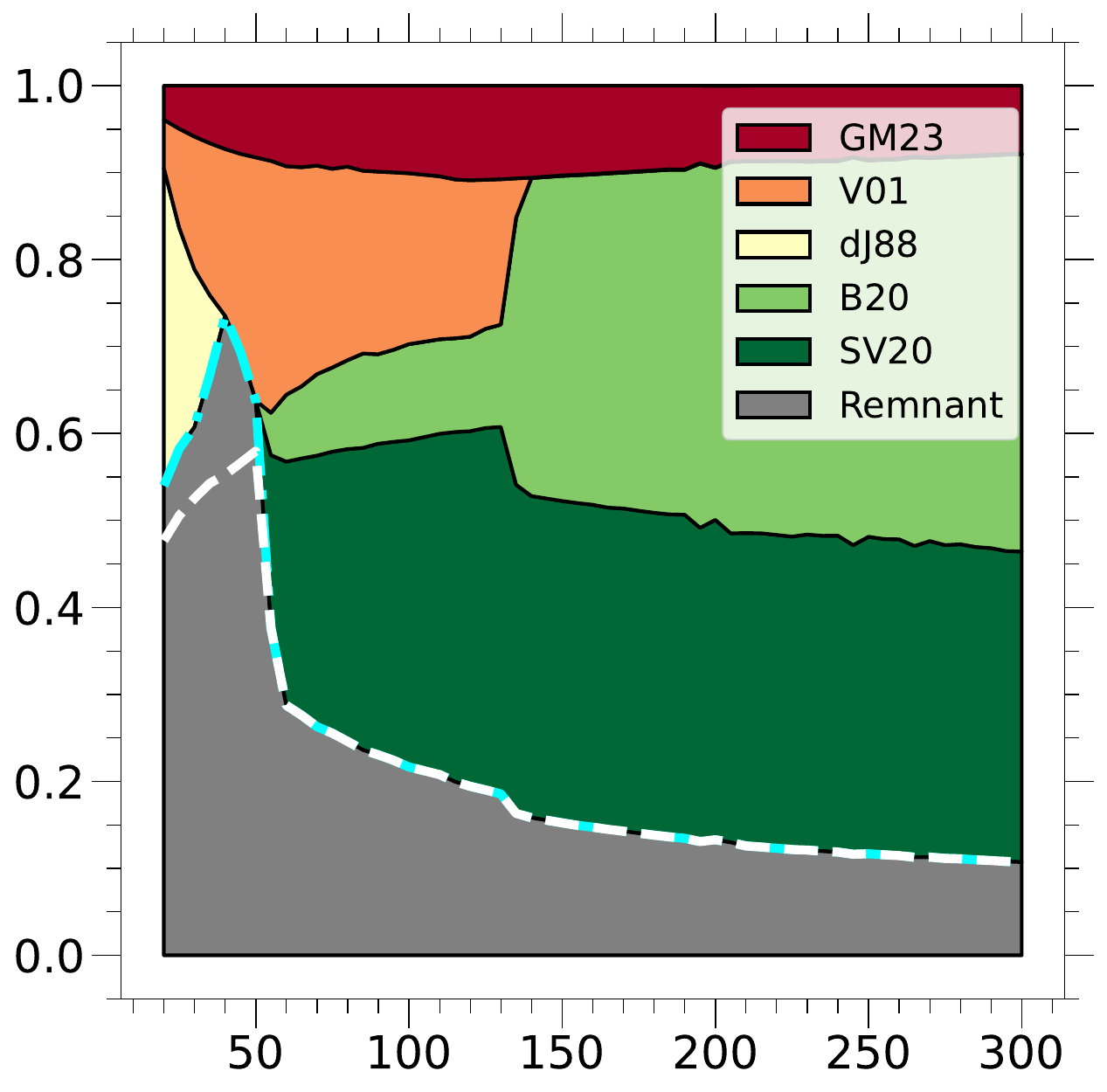}
\includegraphics[width=0.28\textwidth]{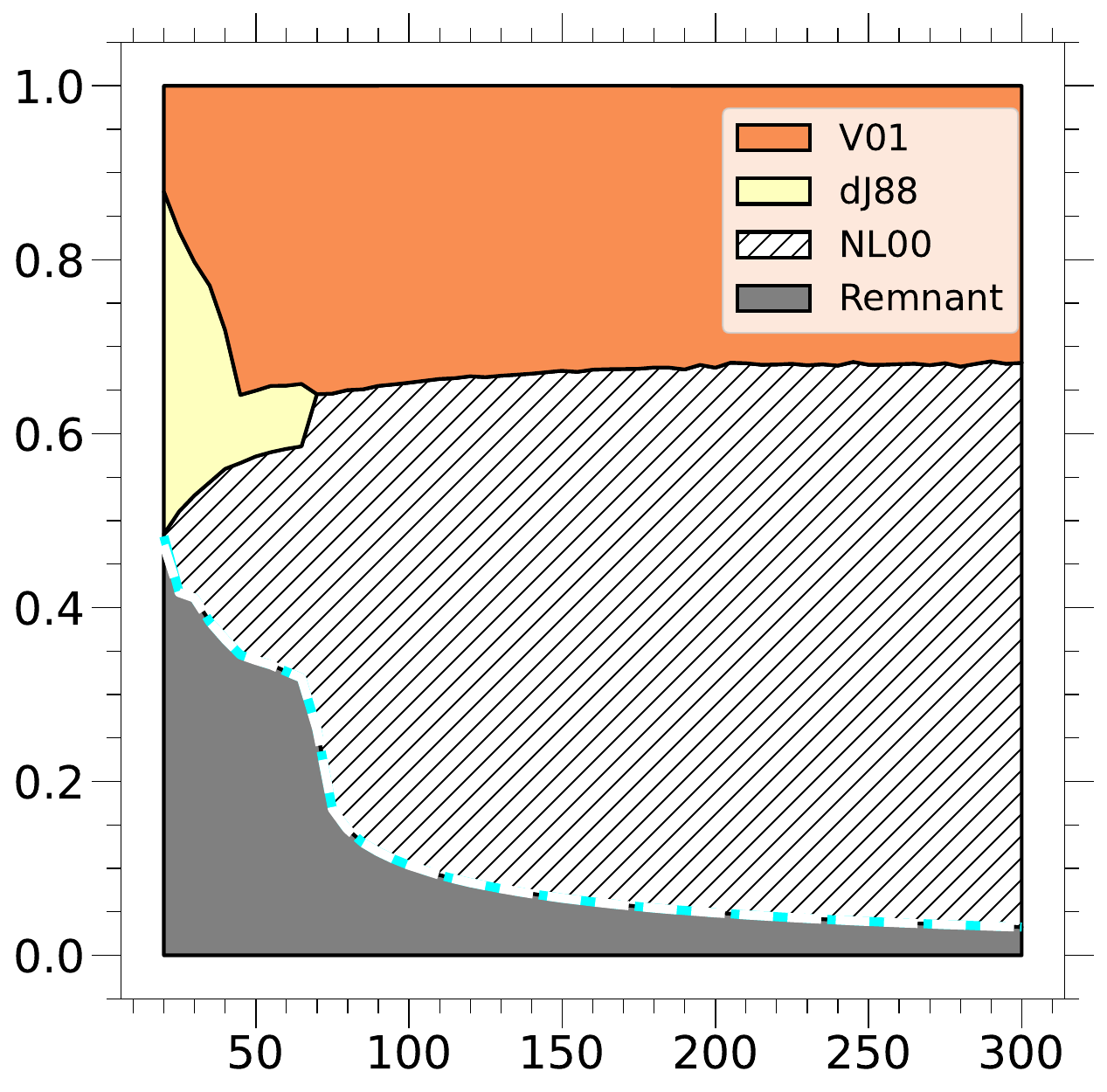}\\
\includegraphics[width=0.30\textwidth]{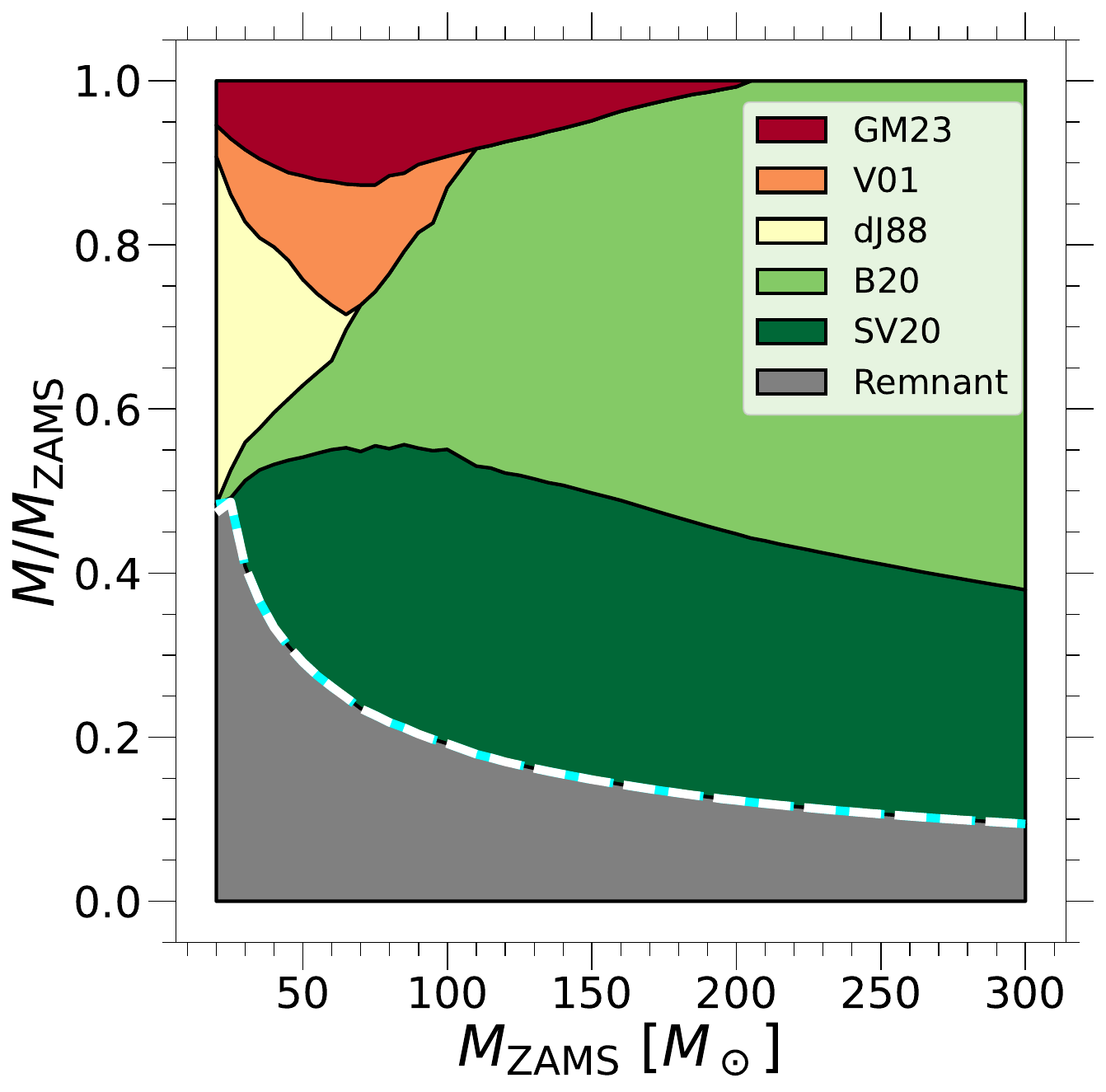}
\includegraphics[width=0.28\textwidth]{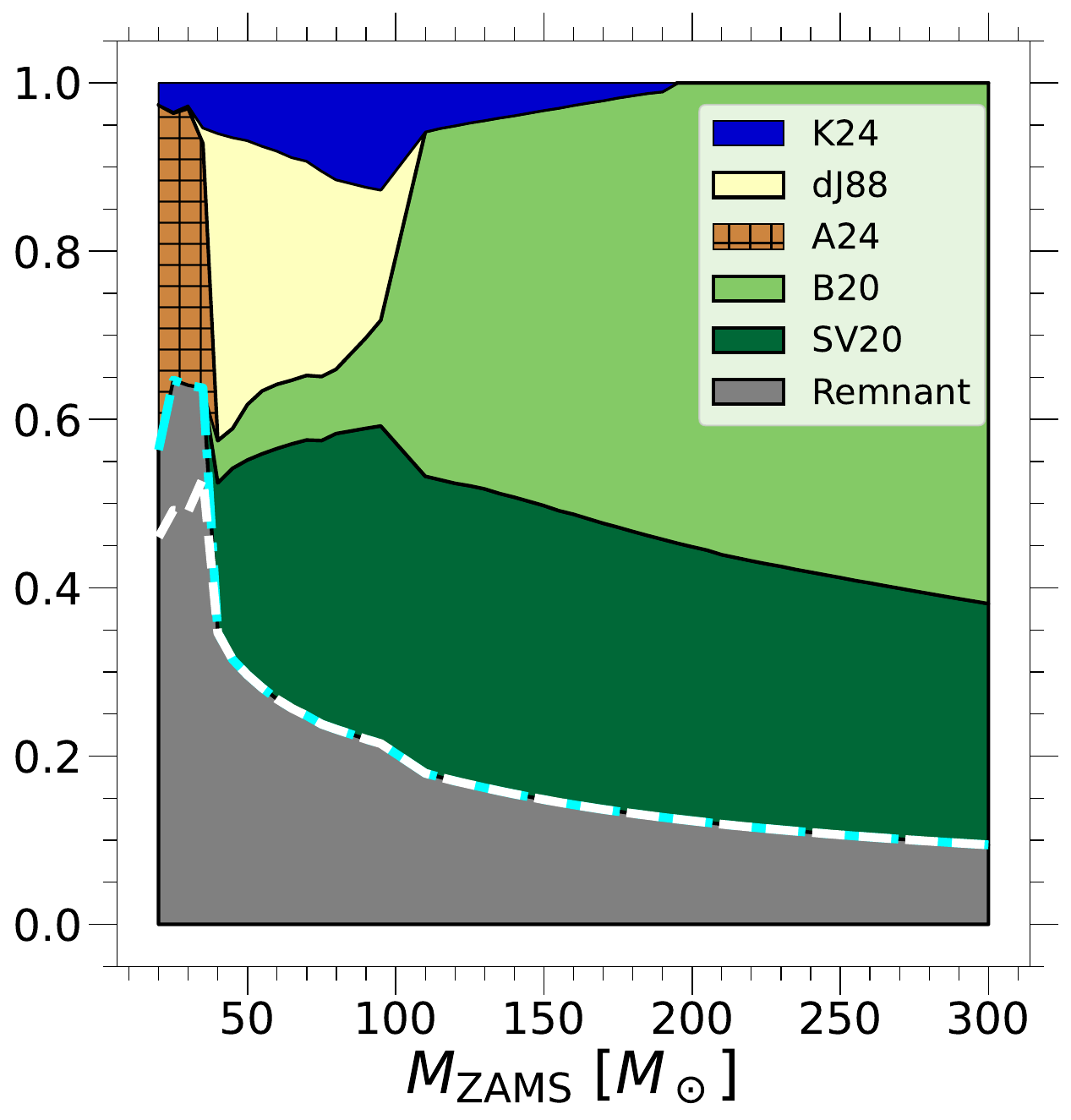}
\includegraphics[width=0.28\textwidth]{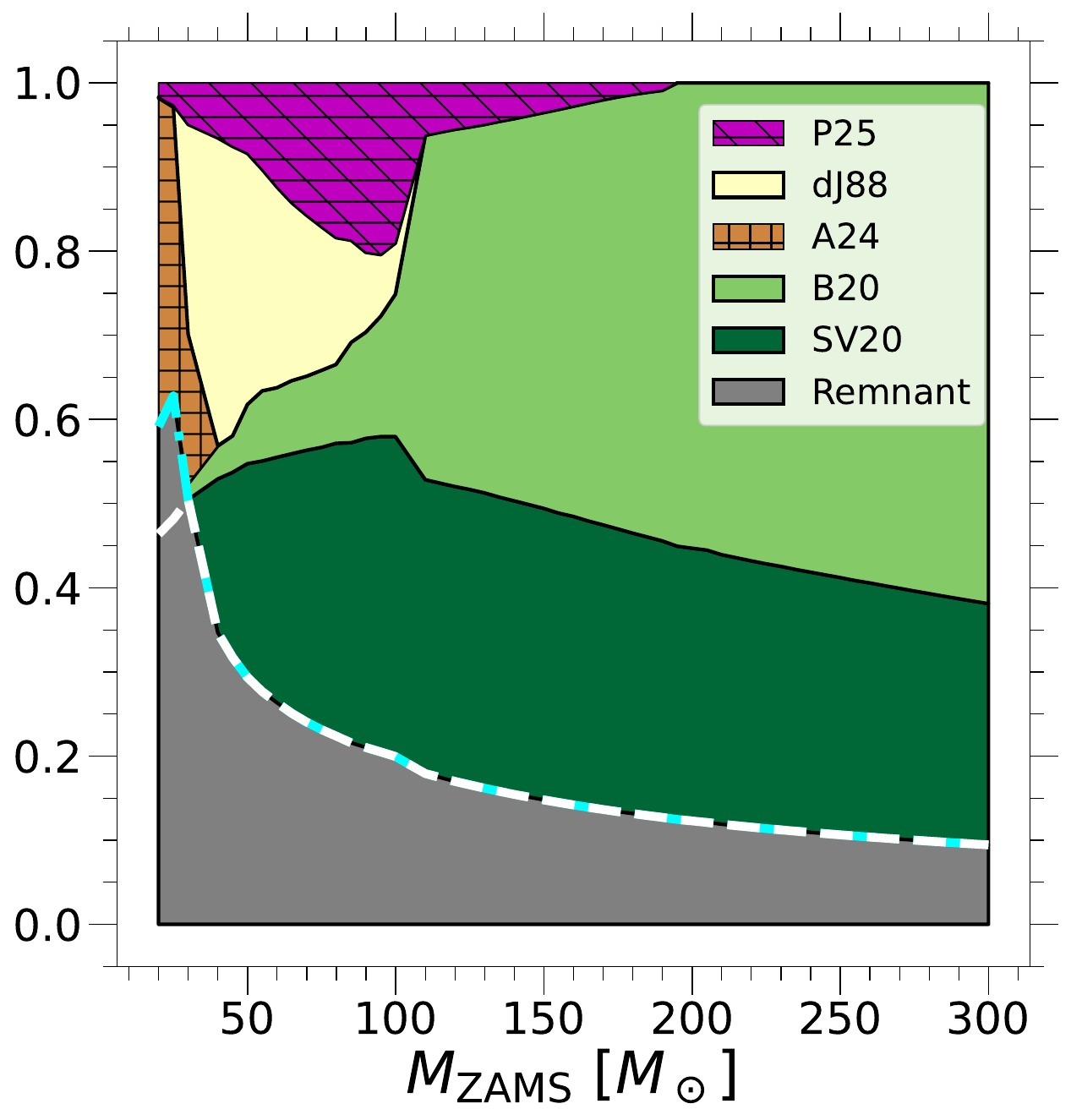}
\caption{Fraction of $M_{\rm ZAMS}$ lost with a specific wind scheme as a function of $M_{\rm ZAMS}$. The gray color is the remnant mass fraction, with the dashed white and dash-dotted aqua lines respectively the final core and envelope mass fractions. Top left: FESc model. Top center: MSc. Top right: $X_{\rm surf}$. Middle left: FESc$_{\rm RSG}$. Center: MSc$_{\rm RSG}$. Middle right: Dutch. Bottom left: FESc$_{\rm highrot}$ model. Bottom center: KABS. Bottom right: Warm\_$\Gamma$. The FESc$_{\rm A24}$ and Dutch-ish models were not included. This visualization follows the format introduced by \cite{Bavera_2023}.}
\label{fig:merge_Mlost}
\end{figure*}

Figure~\ref{fig:merge_Mlost} shows, for each model, the fraction of $M_{\rm ZAMS}$ that was lost for each star from each adopted mass loss scheme. 

\paragraph{\textbf{How cool and expanded a massive star can be?}} A massive star's post-\gls{MS} expansion, and thus its minimum $T_{\rm eff}$, is primarily governed by its mass loss and mixing history. A rough extent to which a star cools down can be inferred on whether it initiates cool supergiant winds (in our case $T_{\rm eff}$\,$\leq$\,10 or 4~kK). The strength of optically thin winds is a key determinant: weaker winds during \gls{MS} allow for a greater degree of post-\gls{MS} expansion to cooler temperatures. This is shown by the KABS and Warm\_$\Gamma$ models, in which stars up to roughly 80~$M_\odot$ at \gls{ZAMS} expand into the cool supergiant wind regime. Conversely, the Dutch model, with its strong V01 winds, predicts the most limited expansion, with only stars up to $\sim$60~$M_\odot$ reaching $T_{\rm eff}$\,$\leq$\,10~kK. Finally, higher initial rotation, as seen in the FESc$_{\rm highrot}$ model, also limits stellar expansion. While this has a negligible effect on the final \gls{BH} mass, it keeps the star more compact and at a higher effective temperature, further inhibiting the onset of cool supergiant winds. 

\paragraph{\textbf{Minor impact from changing optically thick winds transitions on the total mass loss}} Given the same winds scheme, the adoption of different conditions to optically thick winds deeply affects the amount of mass that is lost from a star during its optically thin phase. The FESc models show a diminishing contribution from thin winds as a function of $M_{\rm ZAMS}$, since increasingly higher luminosity levels further anticipate the $\Gamma_{\rm e}$ transition. On the other hand, the MSc models still deeply depend on the mass-loss history to transition to optically thick winds, showing a higher degree of variability in the contribution of thin winds. Finally, the $X_{\rm surf}$\,$\leq$\,0.4 transition to thick winds requires the highest amount of mass loss from the star, since it is limited by the position of the H-burning burning shells. However, despite the stark differences in pre-\gls{WR} mass loss, as also shown in Figure~\ref{fig:BHmass}, the total mass lost is roughly the same. For instance, the $M_{\rm ZAMS}$\,=\,300~$M_\odot$ star varies its $M_{\rm final}$/$M_{\rm ZAMS}$ ratio between roughly only 0.09 (FESc) and 0.11 ($X_{\rm surf}$). This shows that at $Z_\odot$ changing different transitions to optically thick winds has a minor impact on isolated \gls{BH} formation.

\paragraph{\textbf{Envelope conservation until core-collapse}}
As highlighted in the Sections~\ref{sec:results_BH} and ~\ref{subsec:BH_peak_RSG}, restricting cool supergiant winds to \glspl{RSG} leads massive stars to not expel the totality of their envelope and to collapse into \gls{BH}s as supergiants. 
Similarly, even adopting weaker (A24) mass loss for \glspl{RSG} while keeping strong (dJ88) \gls{YSG} mass loss leads to the conservation of stellar envelopes for $M_{\rm ZAMS}\lesssim50~M_\odot$, since \gls{YSG} mass loss alone is not strong enough to shed the H-rich layers of cool supergiants before core-collapse.

\section{Comparison with the literature}
\label{subsec:literature_comparison}

\begin{figure}[!ht]
\centering
\includegraphics[width=0.465\textwidth]{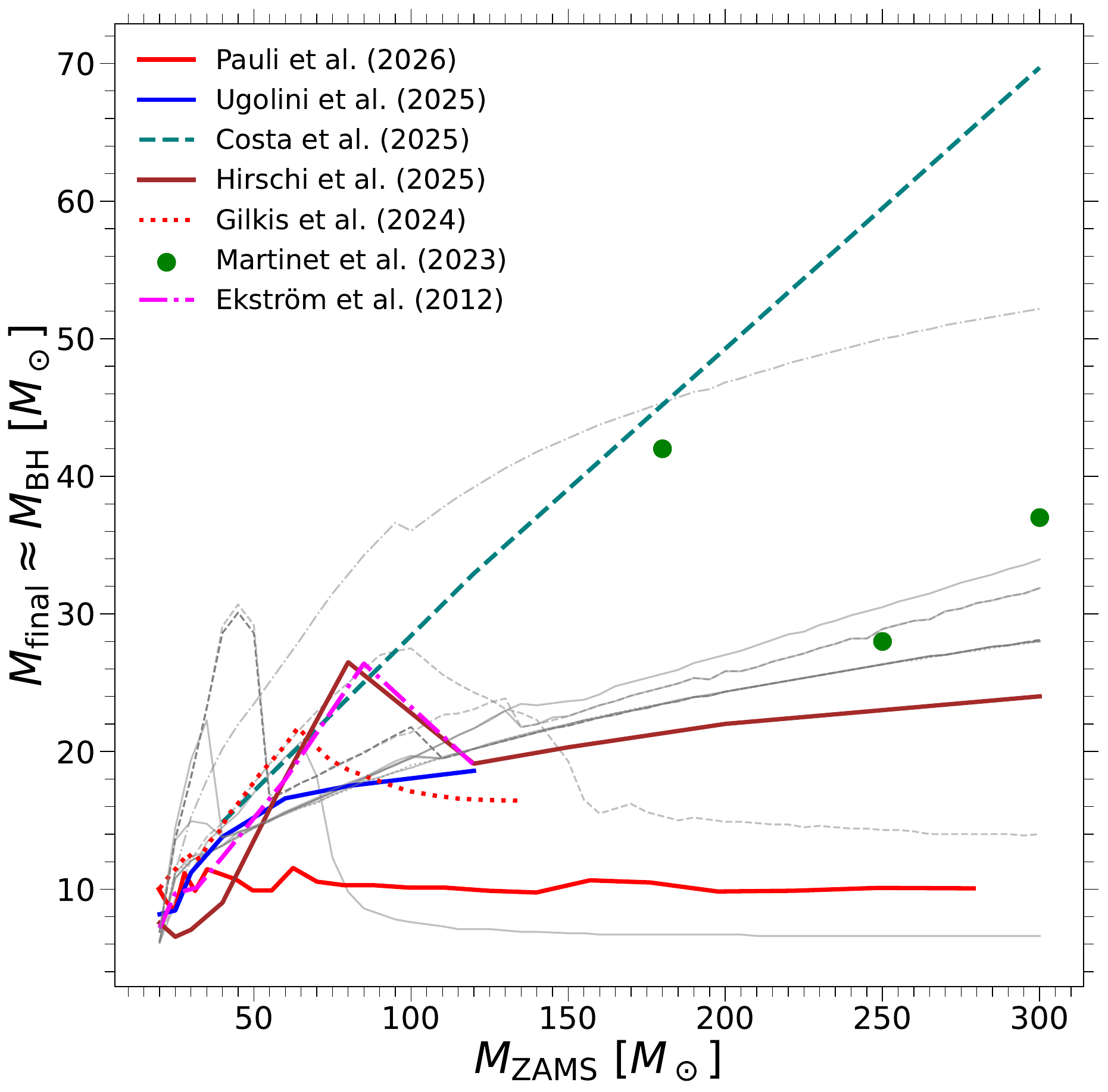}
\caption{Same as Figure~\ref{fig:BHmass}, but with all our models in gray, and the reported $M_{\rm BH}$/$M_{\rm ZAMS}$ ratio from the literature.}
\label{fig:MBH_Liter}
\end{figure}

Figure~\ref{fig:MBH_Liter} shows the recent estimates from the literature for the $M_{\rm BH}$/$M_{\rm ZAMS}$ relation at $Z_\odot$, in comparison with our results.
With the exception of \cite{Ekstrom_2012} that represents one of the most used references for $Z_\odot$ studies, we only included the most recent analyzes that openly report the $M_{\rm BH}$ or final masses for their simulations. In case of multiple models in the same paper, we only report the ones that most closely resemble our initial conditions. From \cite{Gilkis_2024}, \cite{Martinet_2023}, and \cite{Ugolini_2025}, we took their $Z_\odot$ results for $v_{\rm init}= 300$~km\,s$^{-1}$, for \cite{Ekstrom_2012} and \cite{Hirschi_2025}, we took their simulations for $v_{\rm init}/v_{\rm critical}$\,=\,0.4, and lastly we calculated single star models accounting for eruptive mass-loss with the setup from \citet{Pauli_2026} at $Z_\odot$ and an initial rotation velocity of $100~km/s$. We stress that some of these studies only report the stellar mass at core-C depletion. For those, we still apply an upper \gls{BH} mass cap with direct collapse minus 1\% neutrino-driven mass loss.

Most evolutionary models are based on variations of the canonical Dutch winds, with a \gls{BH} mass peak for \gls{VMS}s below 100~$M_\odot$ at \gls{ZAMS}. Most noticeably, the \gls{BH} masses retrieved with the {\tt PARSEC} code by \cite{Costa_2025} seem to increase monotonically as a function of the \gls{ZAMS} mass, entering the \gls{PPSN} regime. This behavior resembles at least qualitatively the one from our FESc$_{\rm noHpoor}$ model, where no dedicated H-free \gls{WR} winds were adopted, resulting results in an underestimate of mass loss.


\cite{Ugolini_2025} is the only study that does not show an evident $M_{\rm BH}$ Dutch peak. The underlying reason is likely the treatment of near-Eddington stellar surfaces in the {\tt FRANEC} code they adopted. In our models, stellar surfaces approaching their Eddington limit transition to optically thick winds, while in its default formulation \citep{Paxton_2011}, {\tt MESA} adopts the super-Eddington mass-loss multiplier from \cite{Paczynski_1986}. In the {\tt FRANEC} code, instead, those layers are automatically ejected from the star \citep{Limongi_2018}. Since we showed in Section~\ref{sec:75Msun} that at $Z_\odot$ this is the mass regime at which many evolutionary tracks experience inflation due to sub-surface near-Eddington stellar regions, this phenomenon can be the reason for the absence of a \gls{BH} mass peak in their models\footnote{This conclusion arose from a private discussion with the main author}.

Finally, \citet{Pauli_2026} yields a near-constant \gls{BH} mass of $M_{\rm BH} \approx 10~M_\odot$ nearly across the whole considered $M_{\rm ZAMS}$ range. This flat relation is explicitly driven by an \gls{LBV} eruptive mass-loss model triggered whenever a star reaches the inflation limit. Within their model framework, stars more massive than $M_{\rm ZAMS} \gtrsim 40\,M_\odot$ experience an eruptive mass-loss event during \gls{MS}, while stars more massive than  $M_{\rm ZAMS} \gtrsim 20\,M_\odot$ undergo these eruptions after shedding their H-rich envelopes during the supergiant phase. Mass is then stripped until the star drops below the inflation threshold for H-rich and H-free envelopes respectively, leading to a pile up of final \gls{WR} masses around $10\,M_\odot$. We emphasize that this mass plateau is strictly dictated by the algorithmic treatment of eruptive mass loss. The calibrated inflation limits for these $Z_\odot$ setups remain untested and may require future readjustments, particularly concerning the \gls{WR} phase that ultimately defines the final remnant mass.

\end{appendix}


\end{document}